\documentclass[a4paper,11pt]{article}
\pdfoutput=1
\usepackage{jheppub} 

\allowdisplaybreaks[4]

\title{\boldmath Double Parton Scattering Effects on the Measurement of the $W$-Boson Mass}
\author[a,b,1]{Rui Zhang\note{Corresponding authors. }}
\author[c,1]{and Zhen Zhang}

\affiliation[a]{Theoretical Physics Division, Institute of High Energy Physics, Chinese Academy of Sciences, 
19B Yuquan Road, Beijing 100049, People's Republic of China}
\affiliation[b]{Shanghai Key Laboratory of Deep Space Exploration Technology, Shanghai Institute of Satellite Engineering, 
3666 Yuanjiang Road, Shanghai 201109, People's Republic of China}
\affiliation[c]{State Key Laboratory of Particle Astrophysics, Institute of High Energy Physics, Chinese Academy of Sciences, 
19B Yuquan Road, Beijing 100049, People's Republic of China}

\emailAdd{zhangr@ihep.ac.cn; zhangzhen@ihep.ac.cn}

\abstract{
Double parton scattering (DPS) corresponds to events where two parton-parton scatterings occur in a single hadron-hadron collision. The DPS effects may arise from the spectator scatterings that are somewhat related to semi-hard QCD activities. In this work, we investigate the DPS effects on the $W$-boson mass measurements. Especially, our analysis reveals that the DPS effects contribute additional missing transverse momenta from spectator scatterings as well as relevant inclusive cross sections, potentially altering the distribution of total missing transverse momenta. Consequently, the DPS effects have the potential to cause an increase in the measured $W$-boson mass by the CDF detector, which helps to understand the deviation of the CDF-II measurements from other measurements and the predicted value in the Standard Model. This could be further validated in upcoming studies, such as the $W$-like analyses of the $Z$ boson production, which will potentially open new avenues for probing QCD dynamics in the semi-hard regime. 
}

\begin{document}
\maketitle
\flushbottom

\section{Introduction}

The QCD factorization theorem~\cite{Collins:1989gx}
claims that, for sufficiently inclusive cross sections, the hadron-hadron collision can be factorized into universal parton distribution functions, convoluted with hard parton-parton scattering cross sections. 
Hence, the hadron-hadron collision can be viewed as a consequence of one parton from each hadron interacting perturbatively. 
However, it is also possible that more than one parton-parton scattering occurs in a single hadron-hadron collision, which is refered to as multi-parton interactions (MPIs)~\cite{Sjostrand:1987su,Landshoff:1975eb,Landshoff:1978fq}. 
Although mostly producing low transverse momentum ($p_{\mathrm{T}}$) particles, the MPI can also generate high $p_{\mathrm{T}}$ particles. 
The former, combined with beam-beam remnants interactions, is known as the underlying event (UE)~\cite{Sjostrand:1985xi,Bengtsson:1986gz,Sjostrand:1987su} activity, which is usually dealt with phenomenological models in the Monte Carlo (MC) event generators~\cite{Sjostrand:2014zea,Bahr:2008pv,Bellm:2015jjp}, whereas the later is called double parton scattering (DPS)~\cite{Gaunt:2009re,Gaunt:2010pi}, depicting events where two pairs of partons participate in strong interactions in a single hadron-hadron collision.

In current data analyses, the MPI effects are considered in the MC simulation codes through the UE and DPS tunes. 
Nonetheless, as per the CMS CP5 tune~\cite{CMS:2019csb}, the UE description does not match the experimental data for leading jets with transverse momenta larger than $5~\mbox{GeV}$. 
This suggests the need for the DPS tune when $p_{\mathrm{T}}$ of leading jets exceeds a threshold of $p_{\mathrm{T},\mbox{\footnotesize cut}}^{\mbox{\scriptsize DPS}}\gtrsim5~\mbox{GeV}$, where $p_{\mathrm{T},\mbox{\footnotesize cut}}^{\mbox{\scriptsize DPS}}$ actually sets a lower limit on $p_{\mathrm{T}}$ of leading jets in the DPS events. 
Notice that all the thresholds used in experiments are higher than 20~GeV~\cite{ATLAS:2013aph,CMS:2013huw,ATLAS:2016rnd}. 
Ergo, neither the UE nor DPS tune effectively describes the events with jet transverse momentum $p_{\mathrm{T}}^j\sim5-20$~GeV, indicating the threshold of $p_{\mathrm{T},\mbox{\footnotesize cut}}^{\mbox{\scriptsize DPS}}\sim5-20$~GeV between the two tunes. 
However, these semi-hard events may have significant influence on measuring inclusive cross sections and missing transverse momenta~\cite{Pumplin:1997ix}. 
Given the high-pile-up environment of the LHC, these effects from DPS are less likely compared to the defunct Tevatron. 
Therefore, it is important to find ways to probe the DPS effects in $\sim5-20$~GeV $p_{\mathrm{T}}$-regions, especially at lower intensity colliders like the Tevatron.

Four years ago, the CDF collaboration reported a new measurement of the $W$-boson mass~\cite{CDF:2022hxs}. 
The measurement is based on the fits to the distributions of the transverse momentum $p_{\mathrm{T}}^\ell$, missing transverse momentum $p_{\mathrm{T}}^{\mbox{\scriptsize miss}}$ and transverse mass $m_{\mathrm{T}}$ for charged leptons $\ell=e^\pm,~\mu^\pm$; the results are~\cite{CDF:2022hxs}: 
\begin{align*}
\\[-7mm]
M_W(p_{\mathrm{T}}^\ell(e))=&80411.4\pm10.7_{\small{\mbox{stat}}}\pm11.8_{\small{\mbox{syst}}}~\mbox{MeV}\,, \nonumber\\
M_W(p_{\mathrm{T}}^{\mbox{\scriptsize miss}}(e))=&80426.3\pm14.5_{\small{\mbox{stat}}}\pm11.7_{\small{\mbox{syst}}}~\mbox{MeV}\,, \nonumber\\
M_W(m_{\mathrm{T}}(e,\nu))=&80429.1\pm10.3_{\small{\mbox{stat}}}\pm8.5_{\small{\mbox{syst}}}~\mbox{MeV}\,, \nonumber\\
M_W(p_{\mathrm{T}}^\ell(\mu))=&80428.2\pm9.6_{\small{\mbox{stat}}}\pm10.3_{\small{\mbox{syst}}}~\mbox{MeV}\,, \nonumber\\
M_W(p_{\mathrm{T}}^{\mbox{\scriptsize miss}}(\mu))=&80428.9\pm13.1_{\small{\mbox{stat}}}\pm10.9_{\small{\mbox{syst}}}~\mbox{MeV}\,, \nonumber\\
M_W(m_{\mathrm{T}}(\mu,\nu))=&80446.1\pm9.2_{\small{\mbox{stat}}}\pm7.3_{\small{\mbox{syst}}}~\mbox{MeV}\,, 
\\[-7mm]
\end{align*}
where $m_{\mathrm{T}}\equiv\sqrt{2\left(p_{\mathrm{T}}^\ell p_{\mathrm{T}}^{\mbox{\scriptsize miss}}-\vec{p}_{\mathrm{T}}^{~\ell} \cdot \vec{p}_{\mathrm{T}}^{~\mbox{\scriptsize miss}}\right)}$ contains the $p_{\mathrm{T}}^{\mbox{\scriptsize miss}}$ information. 
The results from the $p_{\mathrm{T}}^{\mbox{\scriptsize miss}}$ and $m_{\mathrm{T}}$ fits are systematically higher than the $p_{\mathrm{T}}^\ell$ fit. 
With these results from different channels and distributions, they give the combined result
\begin{align*}
\\[-7mm]
M_W=&80433.5\pm6.4_{\small{\mbox{stat}}}\pm6.9_{\small{\mbox{syst}}}~\mbox{MeV}\,, 
\\[-7mm]
\end{align*}
which is not only 6.9$\sigma$ deviation from the standard model (SM) prediction $M_W^{\mbox{\scriptsize SM}}=80359.1\pm5.2~\mbox{MeV}$~\cite{deBlas:2021wap}, but also 3.6$\sigma$ deviation from the most precise result of $M_W=80366.5\pm15.9~\mbox{MeV}$ from other measurements~\cite{ATLAS:2024erm}. 
Similarly, the ATLAS collaboration also reported a higher $W$-boson mass in the $m_{\mathrm{T}}$ fit than in the $p_{\mathrm{T}}^\ell$ fit: 
\begin{align*}
\\[-7mm]
M_W(p_{\mathrm{T}}^\ell)=&80362\pm16~\mbox{MeV}\,, \nonumber\\
M_W(m_{\mathrm{T}})=&80395\pm24~\mbox{MeV}\,. 
\\[-7mm]
\end{align*}
Indeed, we can find similar discrepancies in different situations; see more details in figure~6 of~\cite{ATLAS:2024erm}. To date, the $W$-boson mass anomaly remains an open question in physics.

Given the high level of precision of $\sim10^{-4}$ achieved in the CDF-II measurements, the effects that are normally deemed insignificant may no longer be negligibly small, requiring thorough discussion and examination. One significant feature of these measurements is that most of the signal events involve low-$p_{\mathrm{T}}$ $W$ boson production at the Tevatron. So an accurate simulation of the $W$ boson production associated with soft QCD activities is extremely important to match the experimental uncertainties. In this work, we investigate the effects of DPS on the measurement of the $W$-boson mass at the Tevatron with detector-level simulations, and provide a compelling explanations for the observed $W$-boson anomaly without invoking new physics. Conversely, a detailed analysis of the $W$-mass anomaly allows us to impose extra constraints on phenomenological DPS models.

The rest of the article is organized as follows. In section~\ref{sec:2}, we briefly introduce the DPS effects and present a semi-quantitative analysis of how they modify relevant effective cross sections and kinematic distributions. In section~\ref{sec:3}, combining with the CDF-II data, we detail our simulation methodology, quantify the DPS effects on kinematic distributions, and carry out a further phenomenological analysis. In section~\ref{sec:4}, we discuss some key factors affecting $W$-mass measurements and provide our insights into them. Finally, we summarize our findings and discuss future prospects for probing DPS effects in upcoming measurements in section~\ref{sec:5}.

\section{Double Parton Scattering Effects}
\label{sec:2} 

The DPS effects can potentially produce additional soft hadrons and alter event distributions in low-$p_{\mathrm{T}}$ regions~\cite{Goebel:1979mi}. 
As reported by the CMS (ATLAS) collaboration, same-sign $W^\pm W^\pm$ boson production via DPS is observed at a significance level of 6.2$\sigma$ (8.8$\sigma$), with the DPS effective cross section $\sigma_{\mbox{\footnotesize eff}}=12.2^{+2.9}_{-2.2}~\mbox{mb}$~\cite{CMS:2022pio} ($\sigma_{\mbox{\footnotesize eff}}=10.6\pm1.8~\mbox{mb}$~\cite{ATLAS:2025bcb}). 
Measuring the DPS processes has been done in various channels at the LHC~\cite{LHCb:2012aiv,ATLAS:2013aph,CMS:2013huw,ATLAS:2018zbr,LHCb:2016wuo,ATLAS:2014ofp,LHCb:2015wvu,ATLAS:2016rnd,ATLAS:2016ydt,CMS:2016liw,CMS:2021lxi,LHCb:2023qgu,LHCb:2023ybt}. 
Experiments at the Tevatron also offer consistent DPS effective cross section values with the LHC~\cite{CDF:1993sbj,D0:2014owy,D0:2014vql,D0:2015dyx,D0:2015rpo}. So it is reasonable to set the DPS effective cross section at $12.2~\mbox{mb}$ or $10.6~\mbox{mb}$, in line with the central values measured at the LHC.

In Single Parton Scattering (SPS) analyses, the signal can be contaminated by DPS effects, 
which are essentially composed of a hard SPS process occurring alongside an independent `spectator' parton-parton scattering. 
These spectator scatterings are somewhat associated with soft QCD activities and the resultant radiations can be finally clustered into final-state leptons, jets, and invisible particles.
In current analyses, these final-state leptons are identified with high purity due to their clean signatures, independent of the hadronic activity. 
As a result, their transverse momenta remain largely insensitive to the DPS effects, which primarily influence the jet sector. 
Although soft QCD radiation may contribute non-prompt leptons, it is unlikely to significantly modify the transverse momenta of the hard leptons originating from $W$-boson decays. Furthermore, the final state jets are less likely impacted by soft QCD radiation, since resultant soft hadrons are inherently included in the jet clustering process. Note that the infrared safety of the jet algorithm ensures that the clustered jets can be viewed as contributions from two separate parton-parton scatterings~\cite{Salam:2007xv}. 
Correspondingly, the total missing transverse momentum can also be obtained by the vector sum of the missing transverse momenta from the two constituent parton-parton scatterings. In general, the hadron decays, detector responses, and other relevant factors contribute a small non-zero residue of total transverse momentum, correspondingly leading to extra contributions from spectator scatterings to missing transverse momenta. These extra contributions, related to the DPS processes, will modify kinematic variables such as $p_{\mathrm{T}}^{\mbox{\scriptsize miss}}$ and $m_{\mathrm{T}}$, as well as their distributions.

The DPS cross section can be estimated with the product of the two SPS cross sections to produce processes $S$ and $P$ independently, as~\cite{Humpert:1984ay,Ametller:1985tp}: 
\begin{align}\label{eq:xsection}
\sigma_{S+P}^{\mbox{\scriptsize DPS}}=\frac{n}{2}\frac{\sigma_{S}^{\mbox{\scriptsize SPS}}\sigma_{P}^{\mbox{\scriptsize SPS}}}{\sigma_{\mbox{\scriptsize eff}}}\,, 
\end{align}
in which\footnote{Since the double parton distributions are largely unknown~\cite{Kasemets:2017vyh}, the characterization of DPS effects is grounded in the factorized cross-section ansatz, which has been widely used as an approximation in both theoretical and experimental studies~\cite{Maina:2010vh,DelFabbro:1999tf,Bandurin:2010gn,Maina:2009sj,Kulesza:1999zh,Berger:2009cm}. In fact, the discovery of DPS effects is based on this factorized cross-section formula~\cite{CMS:2022pio,ATLAS:2025bcb}. In this work, we employ this formula to investigate the DPS effects by varying the lower jet $p_{\mathrm{T}}$ cut,  denoted as $p_{\mathrm{T},\mbox{\footnotesize cut}}^{j}$, and requiring the selected jets from the spectator process to satisfy $p^{j}_{\mathrm{T}}\gtrsim p_{\mathrm{T},\mbox{\footnotesize cut}}^{j}$. For decades, it has been extensively validated at $p_{\mathrm{T},\mbox{\footnotesize cut}}^{j}\sim20-30$ GeV~\cite{CDF:1993sbj,ATLAS:2013aph,CMS:2013huw,ATLAS:2016rnd,CMS:2021lxi}. }
$\sigma_{\mbox{\scriptsize eff}}$ is the so-called (universal) effective cross-section, $n$ is the symmetry factor that equals 1 if $S=P$, and 2 otherwise. 
For example, if the $S$ process is defined as a process of dijet production, the DPS cross section reads: 
\begin{align*}
\sigma^{\mbox{\scriptsize DPS}}_{jj+X}=\frac{\sigma_{jj}^{\mbox{\scriptsize SPS}}}{\sigma_{\mbox{\scriptsize eff}}}\times\sigma_{X}^{\mbox{\scriptsize SPS}}\,, 
\end{align*}
where $X\neq jj$ is assumed. 
For a specific SPS process $P~(\to X\neq jj)$, the dijet production process can be viewed as a spectator process, leading to an enhancement of the overall ${\sigma_{jj}^{\mbox{\scriptsize SPS}}/\sigma_{\mbox{\scriptsize eff}}}$ factor for process $P$, seen as a participant process producing $X~(\neq jj)$. 
Note that the MPI can be modeled as the DPS and UE when the jet $p_{\mathrm{T}}$ is above or below a threshold known as the DPS threshold, respectively. The UE effect is already incorporated in common MC simulation codes to describe the low-$p_{\mathrm{T}}$ events. However, the DPS effects should be taken into account as well. Let us denote the DPS threshold as $p_{\mathrm{T},\mbox{\footnotesize cut}}^{\mbox{\scriptsize DPS}}$, which can be determined by comparing simulation to data while scanning the jet transverse momentum cutoff $p_{\mathrm{T},\mbox{\footnotesize cut}}^{j}$. 
Here, we take $p_{\mathrm{T},\mbox{\footnotesize cut}}^{\mbox{\scriptsize DPS}}=10~\mbox{GeV}$ as an example. 
Given that $\sigma_{jj}^{\mbox{\scriptsize SPS}}\sim5\times10^8~\mbox{pb}$ at the 1.96 TeV Tevatron with $p_{\mathrm{T}}^j>p_{\mathrm{T},\mbox{\footnotesize cut}}^{\mbox{\scriptsize DPS}}=10~\mbox{GeV}$, we obtain $\sigma^{\mbox{\scriptsize DPS}}_{jj+X}\sim4\%\times\sigma_{X}^{\mbox{\scriptsize SPS}}$. 
This DPS contribution introduces a $\sim4\%$ correction to all the inclusive processes, consequently modifying the $p_{\mathrm{T}}^{\mbox{\scriptsize miss}}$ and $m_{\mathrm{T}}$ distributions. In the following, we could illustrate these modifications by assigning $X=W^\pm$ for example.

Before our further analysis, please note that the factorized formula~\eqref{eq:xsection} shows a negligible difference in the DPS cross section compared to that using a double-parton distribution function~\cite{Cao:2017bcb}. Furthermore, the formula has been examined by different experiments~\cite{LHCb:2012aiv,ATLAS:2013aph,CMS:2013huw,ATLAS:2018zbr,LHCb:2016wuo,ATLAS:2014ofp,LHCb:2015wvu,ATLAS:2016rnd,ATLAS:2016ydt,CMS:2016liw,CMS:2021lxi,LHCb:2023qgu,LHCb:2023ybt,CDF:1993sbj,D0:2014owy,D0:2014vql,D0:2015dyx,D0:2015rpo}. 
Despite the incomplete theoretical understanding of the DPS effects, the factorized formula~\eqref{eq:xsection} can be used to illustrate how these effects alter the kinematic distributions of events produced in hadron collisions. 
We will employ equation~\eqref{eq:xsection} to quantify how DPS modifies these distributions and biases $M_W$ measurements.

The overall $\sim4\%$ factor is calculated using a DPS threshold of $p_{\mathrm{T},\mbox{\footnotesize cut}}^{\mbox{\scriptsize DPS}}=10~\mbox{GeV}$. 
In fact, the DPS cross section estimation already sets a lower limit of $p_{\mathrm{T},\mbox{\footnotesize cut}}^{\mbox{\scriptsize DPS}}$. 
Generally, $\sigma_{S}^{\mbox{\scriptsize SPS}}$ increases sharply with decreasing $p_{\mathrm{T},\mbox{\footnotesize cut}}^{\mbox{\scriptsize DPS}}$. 
In particular, when $p_{\mathrm{T},\mbox{\footnotesize cut}}^{\mbox{\scriptsize DPS}}$ is small enough, the cross section $\sigma_{S}^{\mbox{\scriptsize SPS}}$, like an inelastic cross section, can even exceed $\sigma_{\mbox{\scriptsize eff}}$; note, the total inelastic cross section, with a vanishing $p_{\mathrm{T}}$ cut, is significantly larger than $\sigma_{\mbox{\scriptsize eff}}$ ~\cite{CMS:2018mlc,CDF:1993wpv}. 
Therefore, equation~\eqref{eq:xsection} is no longer valid for sufficiently small $p_{\mathrm{T},\mbox{\footnotesize cut}}^{\mbox{\scriptsize DPS}}$, and hence the corresponding DPS description; otherwise, $\sigma_{S+P}^{\mbox{\scriptsize DPS}}\gtrsim\sigma_{S}^{\mbox{\scriptsize SPS}}$ or $\sigma_{S+P}^{\mbox{\scriptsize DPS}}\gtrsim\sigma_{P}^{\mbox{\scriptsize SPS}}$, which fails to satisfy our statistical expectations. 
If $p_{\mathrm{T},\mbox{\footnotesize cut}}^{\mbox{\scriptsize DPS}}\sim10~\mbox{GeV}$, the modification factor of $\sim4\%$ closely aligns with the relative uncertainties in the Drell-Yan $W$ and $Z$ boson measurements~\cite{CMS:2020cph,CMS:2014pkt,ATLAS:2016nqi,Fedorko:2006kp}. 
Generally, a decrease in $p_{\mathrm{T},\mbox{\footnotesize cut}}^{\mbox{\scriptsize DPS}}$ leads to a greater modification of the total and differential cross sections. 
Thus, if $p_{\mathrm{T},\mbox{\footnotesize cut}}^{\mbox{\scriptsize DPS}}\lesssim10$ GeV, the effects related to the DPS processes should have been detected in these experiments. Actually, no anomalies have been observed in current measurements. 
Therefore, $p_{\mathrm{T},\mbox{\footnotesize cut}}^{\mbox{\scriptsize DPS}}$ should not be much smaller than 10~GeV. 
Hereafter, we choose $p_{\mathrm{T},\mbox{\footnotesize cut}}^{\mbox{\scriptsize DPS}}=10~\mbox{GeV}$ as a benchmark point.

Next let us illustrate the DPS effects of the spectator dijet production process on the $p_{\mathrm{T}}^{\mbox{\scriptsize miss}}$ distribution of DPS-only events. 
Specifically, the missing transverse momentum $p_{\mathrm{T},\mbox{\scriptsize DPS}}^{\mbox{\scriptsize miss}}$ including the DPS contribution can be described as
\begin{align}
\label{eq:ptmiss}
(\vec p_{\mathrm{T},\mbox{\scriptsize DPS}}^{~\mbox{\scriptsize miss}})^2=(\vec p_{\mathrm{T},\mbox{\scriptsize SPS}}^{~\mbox{\scriptsize miss}})^2+(\vec p_{\mathrm{T},jj}^{~\mbox{\scriptsize miss}})^2+2\vec{p}_{\mathrm{T},\mbox{\scriptsize SPS}}^{~\mbox{\scriptsize miss}}\cdot\vec{p}_{\mathrm{T},jj}^{~\mbox{\scriptsize miss}},  
\end{align}
where the subscript `DPS' refers to the DPS process, and `SPS' is used for the SPS process. 
Roughly, these DPS effects can be divided into two categories: a smearing, broadening the distribution, and a shift, hardening the distribution. 
As shown in equation~\eqref{eq:ptmiss}, the smearing can be induced by the last term, i.e., $2\,\vec{p}_{\mathrm{T},\mbox{\scriptsize SPS}}^{~\mbox{\scriptsize miss}}\cdot\vec{p}_{\mathrm{T},jj}^{~\mbox{\scriptsize miss}}$, where $p_{\mathrm{T},jj}^{\mbox{\scriptsize miss}}$ denotes the missing transverse momentum generated from the spectator dijet production process. 
From our simulation described below, the resolution of the smearing can be estimated as $\langle p_{\mathrm{T},jj}^{\mbox{\scriptsize miss}}\rangle\sim4.2~\mbox{GeV}$, where $\langle p\rangle$ represents the average of $p=|\vec{p}|$ over all events for momentum $\vec{p}$. 
Also, the DPS effects can cause a momentum shift in the $p_{\mathrm{T}}^{\mbox{\scriptsize miss}}$ distribution. 
First, we expand $p_{\mathrm{T},\mbox{\scriptsize DPS}}^{\mbox{\scriptsize miss}}$ in terms of $p_{\mathrm{T},jj}^{\mbox{\scriptsize miss}}$ through equation~\eqref{eq:ptmiss}, followed by averaging by integration of the opening angle between $\vec{p}_{\mathrm{T},\mbox{\scriptsize SPS}}^{~\mbox{\scriptsize miss}}$ and $\vec{p}_{\mathrm{T},jj}^{~\mbox{\scriptsize miss}}$, and consequently, we find that the shift is $\langle\Delta p_{\mathrm{T},\mbox{\scriptsize DPS}}^{\mbox{\scriptsize miss}}\rangle\approx \langle(p_{\mathrm{T},jj}^{\mbox{\scriptsize miss}})^2\rangle/(4\langle p_{\mathrm{T},\mbox{\scriptsize SPS}}^{\mbox{\scriptsize miss}}\rangle)\approx0.11~\mbox{GeV}$ to the order of $(p_{\mathrm{T},jj}^{\mbox{\scriptsize miss}})^2$, where $\Delta p_{\mathrm{T},\mbox{\scriptsize DPS}}^{\mbox{\scriptsize miss}}\equiv p_{\mathrm{T},\mbox{\scriptsize DPS}}^{\mbox{\scriptsize miss}}-p_{\mathrm{T},\mbox{\scriptsize SPS}}^{\mbox{\scriptsize miss}}$. 
Note that the Jacobian peak~\cite{Smith:1983aa} occurs at around $\langle p_{\mathrm{T},\mbox{\scriptsize SPS}}^{\mbox{\scriptsize miss}}\rangle\sim40~\mbox{GeV}$ for the SPS-only process. 
Therefore, the DPS effects are more likely to manifest their existence as a smearing in the $p_{\mathrm{T}}^{\mbox{\scriptsize miss}}$ distribution rather than a shift.

Similar to the $p_{\mathrm{T}}^{\mbox{\scriptsize miss}}$ case, a smearing and a shift can also be induced in the $m_{\mathrm{T}}$ distribution by the DPS effects. In particular, the DPS-induced transverse mass $m_{\mathrm{T}}$ can be estimated as
\begin{align}
\label{eq:mT_DPS}
(m_{\mathrm{T},\mbox{\scriptsize DPS}})^2=(m_{\mathrm{T},\mbox{\scriptsize SPS}})^2+2 p_{\mathrm{T}}^{\ell}\Delta p_{\mathrm{T},\mbox{\scriptsize DPS}}^{\mbox{\scriptsize miss}}-2\vec{p}_{\mathrm{T}}^{~\ell}\cdot\vec{p}_{\mathrm{T},jj}^{~\mbox{\scriptsize miss}}\,. 
\end{align}
Expanding $m_{\mathrm{T},\mbox{\scriptsize DPS}}$ to the second order of $p_{\mathrm{T},jj}^{\mbox{\scriptsize miss}}$ yields
\begin{align*}
\Delta m_{\mathrm{T},\mbox{\scriptsize DPS}}=&\frac{1}{m_{\mathrm{T},\mbox{\scriptsize SPS}}}\left[\frac{p_{\mathrm{T}}^{\ell}}{p_{\mathrm{T},\mbox{\scriptsize SPS}}^{\mbox{\scriptsize miss}}}\vec{p}_{\mathrm{T},\mbox{\scriptsize SPS}}^{~\mbox{\scriptsize miss}}\cdot\vec{p}_{\mathrm{T},jj}^{~\mbox{\scriptsize miss}}-\vec{p}_{\mathrm{T}}^{~\ell}\cdot\vec{p}_{\mathrm{T},jj}^{~\mbox{\scriptsize miss}}\right]\nonumber\\
&+\frac{p_{\mathrm{T}}^{\ell}}{2m_{\mathrm{T},\mbox{\scriptsize SPS}}p_{\mathrm{T},\mbox{\scriptsize SPS}}^{\mbox{\scriptsize miss}}}\left[(p_{\mathrm{T},jj}^{\mbox{\scriptsize miss}})^2-\frac{(\vec{p}_{\mathrm{T},\mbox{\scriptsize SPS}}^{~\mbox{\scriptsize miss}}\cdot\vec{p}_{\mathrm{T},jj}^{~\mbox{\scriptsize miss}})^2}{(p_{\mathrm{T},\mbox{\scriptsize SPS}}^{\mbox{\scriptsize miss}})^2}\right]\nonumber\\
&-\frac{1}{2(m_{\mathrm{T},\mbox{\scriptsize SPS}})^3}\left[\frac{p_{\mathrm{T}}^{\ell}}{p_{\mathrm{T},\mbox{\scriptsize SPS}}^{\mbox{\scriptsize miss}}}\vec{p}_{\mathrm{T},\mbox{\scriptsize SPS}}^{~\mbox{\scriptsize miss}}\cdot\vec{p}_{\mathrm{T},jj}^{~\mbox{\scriptsize miss}}-\vec{p}_{\mathrm{T}}^{~\ell}\cdot\vec{p}_{\mathrm{T},jj}^{~\mbox{\scriptsize miss}}\right]^2\,, 
\end{align*}
where $\Delta m_{\mathrm{T},\mbox{\scriptsize DPS}}\equiv m_{\mathrm{T},\mbox{\scriptsize DPS}}-m_{\mathrm{T},\mbox{\scriptsize SPS}}$. In this case, the smearing stems from the first term, and the shift arises from the last two terms. 
In general, $\vec{p}_{\mathrm{T},\mbox{\scriptsize SPS}}^{~\mbox{\scriptsize miss}}\approx-\vec{p}_{\mathrm{T}}^{~\ell}$ and $p_{\mathrm{T}}^{\ell}\approx p_{\mathrm{T},\mbox{\scriptsize SPS}}^{\mbox{\scriptsize miss}}\approx m_{\mathrm{T},\mbox{\scriptsize SPS}}/2$. Thus it can be observed from our simulation that the smearing has a resolution of $\langle p_{\mathrm{T},jj}^{\mbox{\scriptsize miss}}\rangle\sim4.2~\mbox{GeV}$, akin to that in the $p_{\mathrm{T}}^{\mbox{\scriptsize miss}}$ case. 
In contrast, the $m_{\mathrm{T}}$ shift is approximately zero, i.e., $\langle\Delta m_{\mathrm{T},\mbox{\scriptsize DPS}}\rangle\approx0$, also obtained by integrating the angle out and taking the average. 
Constrained by $\langle m_{\mathrm{T},\mbox{\scriptsize SPS}}\rangle\approx2
\langle p_{\mathrm{T},\mbox{\scriptsize SPS}}^{\mbox{\scriptsize miss}}\rangle$, the relative correction induced by the smearing to the $m_{\mathrm{T}}$ distribution is only half as large as that to the $p_{\mathrm{T}}^{\mbox{\scriptsize miss}}$ distribution, which can be validated using simulations.

\begin{figure}
\centerline{
\includegraphics[width=\columnwidth]{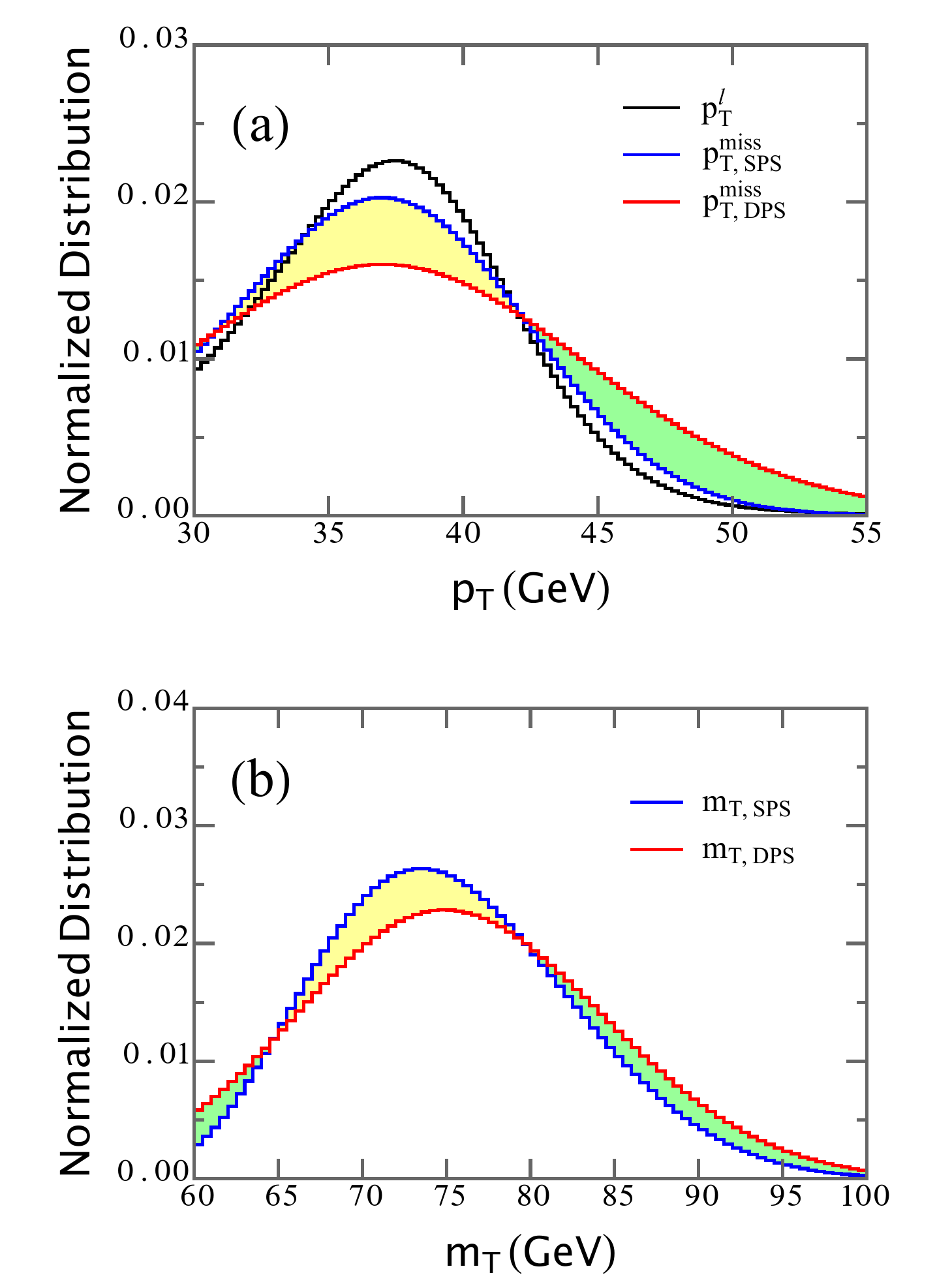}
}
\caption{Normalized distributions for SPS-only and DPS-only contributions. We choose $p_{\mathrm{T}}^j>10~\mbox{GeV}$ for the SPS dijet production process. (a) The black line denotes the $p_{\mathrm{T}}^\ell$ distribution, while blue and red lines represent the $p_{\mathrm{T}}^{\mbox{\scriptsize miss}}$ distributions from the SPS and DPS processes, respectively. (b) The blue and red lines represent the SPS and DPS distributions of $m_{\mathrm{T}}$, respectively. The yellow and green regions illustrate the differences between the DPS and SPS distributions. }
\label{fig:1}
\end{figure}

In short, the DPS effects may smear and shift the distributions of $p_{\mathrm{T}}^{\mbox{\scriptsize miss}}$ and $m_{\mathrm{T}}$. 
These two effects contribute to the measured $M_W$ value from data fitting in different ways: the shift directly increases the measured value, while the smearing tends to harden the distributions above a given cutoff, such as the $p_{\mathrm{T}}^{\mbox{\scriptsize miss}}$ cutoff of $\sim30~\mbox{GeV}$ and the $m_{\mathrm{T}}$ cutoff of $\sim60~\mbox{GeV}$. 
Note that, compared the smearing, the shift has a much weaker influence. 
Hence, the impact on the fitted $m_{W}$ value is primarily attributed to the side effect of smearing, arising from the asymmetry observed in the distributions of $p_{\mathrm{T}}^{\mbox{\scriptsize miss}}$ and $m_{\mathrm{T}}$ over their fitting intervals selected.

In the following analysis, we will focus on the normalized distribution of the signal events from the $W$ boson production processes. 
Note that the main background events originate from the $Z$ boson production processes, only making up $\sim7\%$ of all the signal events~\cite{CDF:2022hxs}. 
Although the distribution of the missing transverse momenta from the background $Z$ boson production processes may be affected by the DPS effects, we neglect the background events from the $Z$ boson production processes due to their small contribution to total number of events.

\section{Simulation and Analysis}
\label{sec:3} 

To illustrate the DPS effects, we perform detailed simulations at the 1.96 TeV Tevatron. 
Overall, we generate 100 million dijet production events using {\tt MadGraph5}~\cite{Alwall:2014hca} at leading order with {\tt CT18NNLO} parton distribution functions~\cite{Hou:2019qau} at parton level. In particular, we apply the cut $p_{\mathrm{T}}^j>10~\mbox{GeV}$ at the parton level. 
Then, we utilize {\tt PYTHIA}~\cite{Sjostrand:2014zea} for parton shower and hadronization simulation and {\tt Delphes}~\cite{deFavereau:2013fsa} for detector simulation. Consequently, we obtain $\sigma_{jj}^{\mbox{\scriptsize SPS}}=4.6\times10^8~\mbox{pb}$ and $\sigma_{jj}^{\mbox{\scriptsize SPS}}/\sigma_{\mbox{\scriptsize eff}}=3.7\%-4.3\%\sim4\%$.

We obtain the $p_{\mathrm{T},jj}^{\mbox{\scriptsize miss}}$ distribution from the simulated dijet events and extract the $p_{\mathrm{T},\mbox{\scriptsize SPS}}^{\mbox{\scriptsize miss}}$ and $m_{\mathrm{T},\mbox{\scriptsize SPS}}$ distributions from the CDF-II data. 
Then we calculate the distributions of $p_{\mathrm{T},\mbox{\scriptsize DPS}}^{\mbox{\scriptsize miss}}$ and $m_{\mathrm{T},\mbox{\scriptsize DPS}}$ for the DPS-only processes; more calculation details are shown in appendix~\ref{appendix:A}. 
In panels (a) and (b) of figure~\ref{fig:1}, we illustrate the normalized distributions of SPS-only and DPS-only events, respectively. 
In comparison to the SPS distributions, the DPS distributions peak at nearly the same locations, indicating that the shift effect is insignificant, but feature a significant extra smearing effect. After including the smearing effect, the $p_{\mathrm T}^{\mbox{\scriptsize miss}}$ and $m_{\mathrm T}$ distributions are both hardened a lot compared to those from the SPS-only process, resulting in an obvious increase in events in the high-energy band of $p_{\mathrm T}^{\mbox{\scriptsize miss}}\sim43-55$~GeV or $m_{\mathrm T}\sim80-100$~GeV. When fitting, we usually limit the range to be above a certain cutoff like $p_{\mathrm T}^{\mbox{\scriptsize miss}}\sim30$~GeV or $m_{\mathrm T}\sim60$~GeV. The increase in the number of high-energy events has increased the fitting weight of data points in that energy range; see appendix~\ref{appendix:A} for details. As a result, the fitted $M_W$ will become larger.

Now we quantitatively estimate the smearing effect. For the $p_{\mathrm{T}}^{\mbox{\scriptsize miss}}$ fit, the DPS-only distribution deviates significantly from the SPS-only distribution, with $\sim12\%$ of events being moved from the yellow area to the green area, as illustrated in panel (a) of figure~\ref{fig:1}. 
The yellow area is centered at $p_{\mathrm{T}}^{\mbox{\scriptsize miss}}\approx37~\mbox{GeV}$, and the green area is at $p_{\mathrm{T}}^{\mbox{\scriptsize miss}}\approx47~\mbox{GeV}$. 
This corresponds to an average increase of $\sim1.2~\mbox{GeV}$ in $p_{\mathrm{T}}^{\mbox{\scriptsize miss}}$. 
Here, we estimate the DPS effects solely based on normalized $p_{\mathrm{T}}^{\mbox{\scriptsize miss}}$ distributions. 
Note that the DPS events only constitute $\sim4\%$ of total events. 
Consequently, after including all the SPS and DPS events, the average increase in $p_{\mathrm{T}}^{\mbox{\scriptsize miss}}$ is $\sim48~\mbox{MeV}$. Roughly, one has $\langle p_{\mathrm{T}}^{\mbox{\scriptsize miss}}\rangle\sim M_W/2$. 
Therefore, we expect a $W$-boson mass shift of $\sim90~\mbox{MeV}$ in the $p_{\mathrm{T}}$ fit. 
Also for the $m_{\mathrm{T}}$ fit, $\sim4.4\%$ of events transition from the upper portion of the SPS distribution at $m_{\mathrm{T}}\sim72~\mbox{GeV}$ to the right wing at $m_{\mathrm{T}}\sim88~\mbox{GeV}$, while $1.7\%$ of events move to the left wing at $m_{\mathrm{T}}\sim62~\mbox{GeV}$. 
Given $\langle m_{\mathrm{T}}\rangle\sim M_W$, we finally expect a $W$-boson mass shift of $\sim25~\mbox{MeV}$ in the $m_{\mathrm{T}}$ fit. 
The $W$-boson mass shift might also be subject to modifications due to the statistic uncertainties and systematic uncertainties in the $p_{\mathrm{T}}^{\mbox{\scriptsize miss}}$ and $m_{\mathrm{T}}$ distributions. 
Considering all these factors, including shift, smearing, and uncertainties, we anticipate a mass shift of $\sim100~\mbox{MeV}$ in the $p_{\mathrm{T}}^{\mbox{\scriptsize miss}}$ fit and $\sim30~\mbox{MeV}$ in the $m_{\mathrm{T}}$ fit. 
Given the substantial uncertainty of $\sim20~\mbox{MeV}$ in each fit~\cite{CDF:2022hxs}, an apparent agreement between the two fitted values may arise from the common practice of optimizing analysis parameters to achieve their expected consistency in preliminary analyses, which can unintentionally introduce biases that compel different measurements to converge~\cite{Roodman:2003rw,Klein:2005di}. In order to maximally suppress these biases, it is currently preferable to evaluate the mass shift using their combined result.

In each fit, the number of signal events is the summation of the SPS-only and DPS-only events, namely
\begin{align}
n^{\mbox{\scriptsize SM}}=n_{\mbox{\scriptsize SPS}}^{\mbox{\scriptsize SM}}+n_{\mbox{\scriptsize DPS}}^{\mbox{\scriptsize SM}}\,. 
\end{align}
The total sample size is rescaled back to the sample size of $n=4236186$ used by the CDF collaboration, and the combination of the SPS and DPS distributions is based on the ratio $n_{\mbox{\scriptsize DPS}}/n_{\mbox{\scriptsize SPS}}=\sigma_{\mbox{\scriptsize DPS}}/\sigma_{\mbox{\scriptsize SPS}}$. 
We divide the events into 100 bins with $p_{\mathrm{T}}^{\mbox{\scriptsize miss}}$ and 80 bins with $m_{\mathrm{T}}$. 
The $p_{\mathrm{T}}^{\mbox{\scriptsize miss}}$ fit is performed in the region with $32~\mbox{GeV}<p_{\mathrm T}^{\mbox{\scriptsize miss}}<48~\mbox{GeV}$, while the $m_{\mathrm{T}}$ fit in the region with $65~\mbox{GeV}<m_{\mathrm T}<90~\mbox{GeV}$. 
Following the CDF collaboration, for a given $M_W$ value, we use the SPS templates (see appendix~\ref{appendix:A}) to fit the $p_{\mathrm{T}}^{\mbox{\scriptsize miss}}$ and $m_{\mathrm{T}}$ distributions of signal events. 
And the goodness of fit is calculated as
\begin{align}
\chi^2=\sum\limits_{i}\frac{(n_i-n_i^{\mbox{\scriptsize SM}})^2}{n_i^{\mbox{\scriptsize SM}}+(f_{\mbox{\scriptsize syst}} n_i^{\mbox{\scriptsize SM}})^2}\,, 
\end{align}
where $n_i$ is the event number of the $i$-th bin, and  $f_{\mbox{\scriptsize syst}}$ is adapted to be $10\%$.

\begin{figure}
\centerline{
\includegraphics[width=\columnwidth]{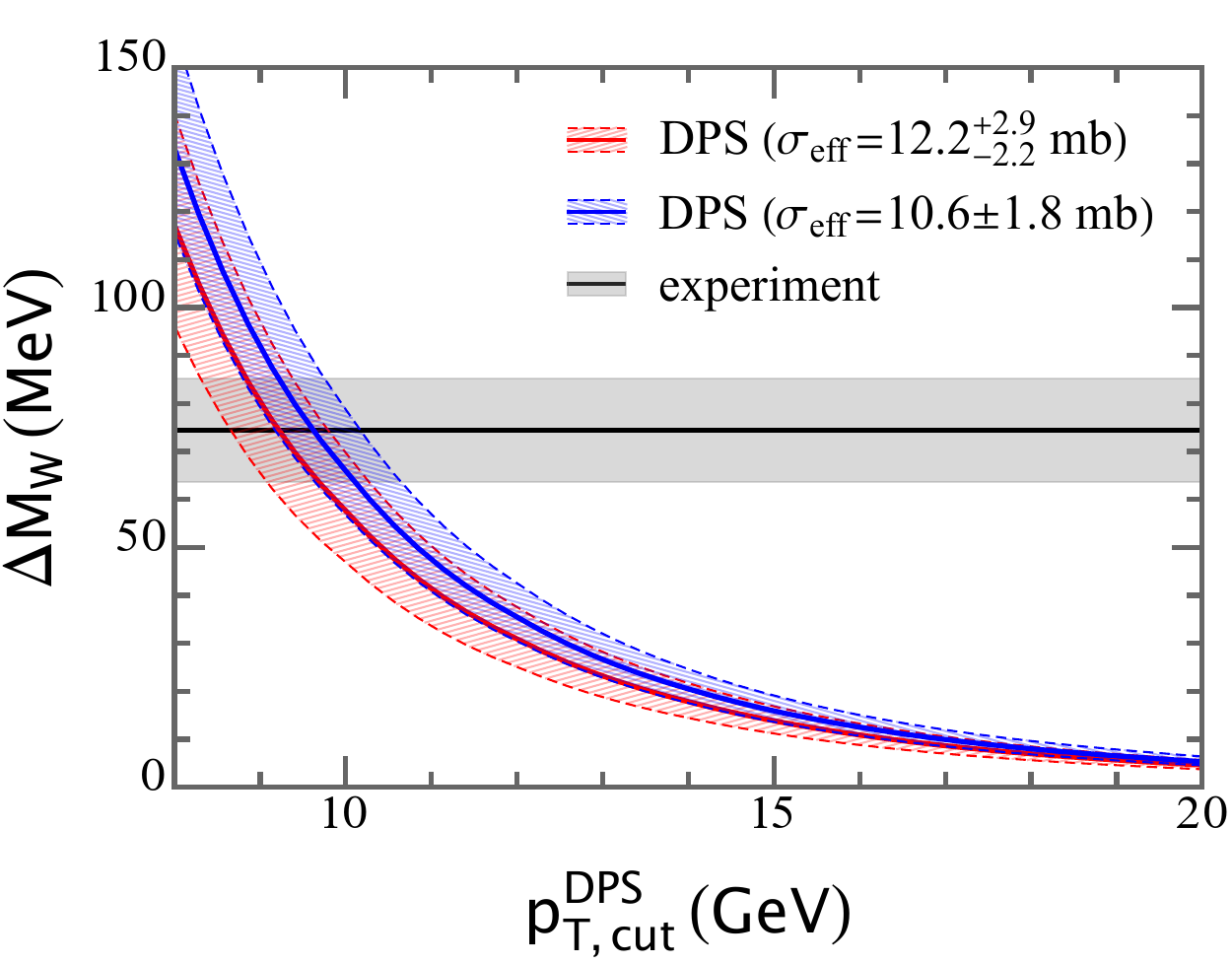}
}
\caption{The mass shift $\Delta M_W$ as a function of the DPS threshold $p_{\mathrm{T},\mbox{\footnotesize cut}}^{\mbox{\footnotesize DPS}}$, extracted from the CDF-II data. 
The region shaded with red (blue) lines displays the measured $W$-boson mass shift due to the DPS effects for $\sigma_{\mbox{\footnotesize eff}}=12.2^{+2.9}_{-2.2}~\mbox{mb}$~\cite{CMS:2022pio} ($\sigma_{\mbox{\footnotesize eff}}=10.6\pm1.8~\mbox{mb}$~\cite{ATLAS:2025bcb}). 
The upper and lower boundaries of the red (blue) region correspond to the values of $\sigma_{\mbox{\footnotesize eff}}$ at its $1\sigma$ lower and upper limits. 
The gray region indicates the discrepancy between the CDF-II result and the SM prediction; the solid black line shows the central value, while the gray band corresponds to the $1\sigma$ confidence interval, incorporating both experimental and theoretical uncertainties. 
}
\label{fig:2}
\end{figure}

For various DPS thresholds $p_{\mathrm{T},\mbox{\footnotesize cut}}^{\mbox{\scriptsize DPS}}$, we can apply the previous procedure and obtain the fitted values of $M_W$, denoted by $M_W\equiv M_W^{\mbox{\scriptsize SM}}+\Delta M_W$, from the $p_{\mathrm{T}}^{\ell}$, $p_{\mathrm{T}}^{\mbox{\scriptsize miss}}$ and $m_{\mathrm{T}}$ distributions. Note that $p_{\mathrm{T},\mbox{\footnotesize\rm cut}}^{\mbox{\scriptsize\rm DPS}}$ serves as a lower limit on the transverse momenta of leading jets in the spectator processes, and it is directly related with the strength of the DPS effects. 
Then, we can combine the fitting results from the three distributions with a weighted average by assuming that the uncertainty of $M_W$ in each fit is the same as that of the CDF-II measurement~\cite{deBlas:2021wap}. 
Accordingly, we can derive the combined values of $M_W$ or the mass shift $\Delta M_W$. As expected, $\Delta M_W$ will decrease with the increasing threshold $p_{\mathrm{T},\mbox{\footnotesize cut}}^{\mbox{\scriptsize DPS}}$, due to the decreasing dijet cross section. 
Actually, it can be roughly expressed as 
\begin{align}
\label{MWfit0}
\Delta M_W=\Delta_{10}~\,(p_{\mathrm{T},\mbox{\footnotesize cut}}^{\mbox{\footnotesize DPS}}/10~\mbox{GeV})^{-3}\,, 
\end{align}
where $\Delta_{10}=56_{-9}^{+10}$ MeV for CMS's $\sigma_{\mbox{\footnotesize eff}}=12.2^{+2.9}_{-2.2}~\mbox{mb}$~\cite{CMS:2022pio} or $\Delta_{10}=62_{-8}^{+10}$ MeV for ATLAS's $\sigma_{\mbox{\footnotesize eff}}=10.6\pm1.8~\mbox{mb}$~\cite{ATLAS:2025bcb}. 
Here, the uncertainty in $\Delta M_W$ is derived from that in $\sigma_{\mbox{\scriptsize eff}}$. 
Note that the mass shift $\Delta M_W$ is rather significant. In fact, even for a large threshold of $p_{\mathrm{T},\mbox{\footnotesize cut}}^{\mbox{\scriptsize DPS}}\sim20~\mbox{GeV}$, it is already comparable to other higher-order QCD effects, such as the N$^3$LL+NNLO correction of $\sim10~\mbox{MeV}$ to the $W$-boson mass~\cite{Isaacson:2022rts}. 
Specifically in figure~\ref{fig:2}, the scale dependence of the mass shift $\Delta M_W$ is shown with the red and blue lines for the two typical values of $\Delta_{10}$, respectively. 
Correspondingly, we also indicate the deviation of the measured CDF-II $M_W$ value from the SM prediction with a black line, along with its $1\sigma$ uncertainty band, which combines both the experimental and SPS-related theoretical uncertainties in the $W$ boson production. 
Especially in the intersection region indicated in figure~\ref{fig:2}, where the DPS threshold is $p_{\mathrm{T},\mbox{\footnotesize cut}}^{\mbox{\footnotesize DPS}}\sim10~\mbox{GeV}$, the mass shift due to the DPS effects can account well for the $W$-boson mass discrepancy.

The DPS effects can also affect the $Z$ boson production. Although the $Z$-boson mass is typically measured in $Z\to\ell\ell$ events using the dilepton invariant mass, which is largely unaffected by the DPS effects, we can alternatively perform the measurement using so-called $W$-like events~\cite{CMS:2016nnd}. These correspond to the $Z\to\ell\ell$ decays in which one of the two leptons is not observed, mimicking the signature of $W\to\ell\nu$. 
Indeed, the $W$-like event sample can be derived from the $Z\to\ell\ell$ background in the $M_W$ measurements, with one of the two leptons fails reconstruction. 
In fact, combining simulations and the CDF-II data~\cite{CDF:2022hxs}, we extract the kinematic distributions of these $W$-like events, including both SPS-only and DPS-only events, and study the DPS effects on the $W$-like measurement of the $Z$-boson mass; see appendix~\ref{appendix:B} for details. 
Like the $W$-boson mass, the measured $Z$-boson mass should be observed to decrease with the DPS threshold $p_{\mathrm{T},\mbox{\footnotesize cut}}^{\mbox{\footnotesize DPS}}$. Denote the observed $Z$-boson mass by $M_Z\equiv M_Z^{\mbox{\scriptsize SM}}+\Delta M_Z$, where $M_Z^{\mbox{\scriptsize{SM}}}=91204.7~\mbox{MeV}$ is obtained from the global fit to precision electroweak data within the SM. Then, we find a mass shift $\Delta M_Z\sim85~\mbox{MeV}$ when the DPS threshold is set to $p_{\mathrm{T},\mbox{\footnotesize cut}}^{\mbox{\footnotesize DPS}}\sim$10~GeV; see appendix~\ref{appendix:B} for details. Interestingly, the mass shift $\Delta M_Z$ decreases with the DPS threshold $p_{\mathrm{T},\mbox{\footnotesize cut}}^{\mbox{\footnotesize DPS}}$ in the same power-law pattern as that observed for the $W$-boson mass. 

If the observed $M_W$ discrepancy solely stems from new physics beyond the SM, such a universal effect cannot account for the divergence between LHC and Tevatron measurements. Instead, this tension likely reflects potential differences between the two colliders in terms of experimental analyses or collider environments. Specifically, it may result from certain effects overlooked in current analyses, notably the varying influence of DPS effects in distinct pile-up environments. This can be tested by searching for a corresponding shift in the measured $Z$-boson mass using a $W$-like analysis at the Tevatron.

\section{Further Insights into $M_W$ Measurements}
\label{sec:4}

Precise measurements of the $W$-boson mass rely on an accurate description of the missing transverse energy (MET). Hadronic calibration is essential for $W$-boson mass measurements, as it corrects the detector's energy scale and resolution, thereby improving the MET reconstruction. Note that, to a first approximation, this calibration can be modeled as a function of the energy scale and resolution~\cite{CMS:2011shu}. It aims to addresses the residual discrepancies between measured data and detector-level simulations in samples such as $Z$ boson production, using precise world-average values, like the $Z$-boson mass, as a reference standard to constrain these simulations. Here, these average values are typically obtained from global fits to precision electroweak data within the SM. For example, when performing calibrations for the track momentum and calorimeter energy measurements, the $Z$ events are selected to contain exactly either two opposite-sign muons or electrons, satisfying standard tight isolation and identification requirements~\cite{CMS:2012nsv,CMS:2015xaf}, and then, the final-state kinematics are adjusted by imposing constraints such that the reconstructed invariant mass $m_{\ell\ell}$ of the lepton pair to match the precise world-average $Z$-boson mass. Subsequently, this adjustment can induce changes in the reconstructed properties of visible particles, including their energy scale and energy resolution. Each of these changes can be universally described by a corresponding calibration factor derived from addressing the residual discrepancies~\cite{CMS:2016lmd}, capturing the instrumental behavior regardless of the specific event topology. However, the calibration factor, defined as a function of particle flavor, pseudo-rapidity, and $p_{\mathrm{T}}$, varies with the pile-up environment and the jet algorithm employed~\cite{CMS:2016lmd}. Specifically, in high-pile-up environments, the particle's neighborhood is susceptible to significant contamination from pile-up and the UE. The large pile-up results in a relatively uniform background in pseudo-rapidity and azimuth~\cite{Cacciari:2008gn}, which can be corrected using a calibration factor derived from techniques based on jet areas~\cite{Cacciari:2007fd}. In contrast, applying such methods in low-pile-up environments is considerably more challenging, as the background lacks this uniformity and is dominated by stochastic noise. 

When measuring the $W$-boson mass, the CDF collaboration uses the invariant mass $m_{\ell\ell}$ and transverse momentum $p_{\mathrm{T}}^{\ell\ell}$ of the lepton pair in $Z\to\ell\ell$ events to calibrate the kinematic measurements of leptons and hadrons~\cite{CDF:2022hxs}. As a result, the transverse kinematic quantities of visible particles, such as the lepton transverse momentum $p_{\mathrm{T}}^{\ell}$, are scaled using the derived calibration factors. 
Since the hadronic distributions in $Z$-boson events are not compared to standard hadronic templates, the calibration factors fail to integrate the DPS effects in both the hadronic observables and the reconstructed recoil. Consequently, the $Z$-based calibration itself cannot eliminate the DPS contributions. When these calibration factors are applied to the hadronic variables in $W$-boson events, the DPS effects remain unaccounted for, inevitably introducing DPS-induced biases into $W$-boson analyses. Subsequently, the derived hadronic recoil has not been corrected to reduce the DPS effects. In fact, we can measure the hadronic recoil $\vec{u}_{\mathrm{T}}$, including both the parallel ($u_\parallel$) and perpendicular ($u_\perp$) parts along the transverse momentum $\vec{p}_{\mathrm{T}}^{~V}$ of the vector-boson $V$, directly from the uncorrected transverse momenta of visible hadrons, and then construct the corresponding hadronic distributions. However, these hadronic observables and their distributions, referred to as ``directly measured hadronic information'', which are sensitive to DPS-induced kinematic effects\footnote{Note that the hadronic recoil comprises a mixture of hadronic deposits. However, relying solely on the total hadronic recoil inherently obscures its internal composition, making it impossible to disentangle whether the observed activity stems from isotropic DPS radiation or from detector-level energy losses associated with initial-state radiation. These distinct physical contributions can only be resolved through detailed hadronic distributions~\cite{Menke:2024tjk}. While such hadronic information is accessible when measuring the recoil directly for a target process, it cannot be retained when transferring a calibration derived from one process to another, as the coarse-grained mapping discards event-level topological details. }, have not been used as inputs for modeling hadronic activity in the CDF measurement~\cite{CDF:2022hxs} of the $W$-boson mass, potentially introducing residual biases in the hadronic recoil calibration.

To suppress the QCD radiations, the hadronic calibration should include the ``directly measured hadronic information'' from $Z\to\ell\ell$ events. Furthermore, deep neural-network techniques like DeepMET~\cite{Feng:2020iyy,CMS:2025prt} can be employed to optimize the $p_{\mathrm{T}}^{\mbox{\scriptsize miss}}$ reconstruction. Indeed, these methods have already been implemented in the ATLAS and CMS analyses~\cite{CMS:2024lrd,ATLAS:2024erm}. In general, the calibration using the $Z\to\ell\ell$ data imposes much tighter constraints on leptonic observables, such as $m_{\ell\ell}$ and $p_{\mathrm{T}}^{\ell\ell}$, than on hadronic observables, such as the hadronic recoil $\vec{u}_{\mathrm{T}}$. In fact, leptonic observables have the lowest uncertainties and carry the highest weight  due to their high precision and direct constraint from the high-precision $Z$-boson invariant mass value, thus dominating the calibration procedure. In comparison, the hadronic observables are subject to larger uncertainties and lack direct constraint from the $Z$-boson mass. In fact, as the whole procedure is predominantly driven by high-precision leptonic kinematics, the hadronic calibration contributes a total uncertainty of only $\sim4$~MeV to the ATLAS $M_W$ measurement~\cite{Davies:2014rdz,ATLAS:2009zsq}.

Furthermore, the DPS effects have a negligible influence on final state leptons. 
In other words, lepton information is insensitive to the DPS effects. 
Specifically in DPS events, the resultant QCD radiations, comprising quarks and gluons, produces energetic hadrons that are detectable, as well as low-energy hadrons and other particles that can not been seen in detectors. They are grouped into jets or contribute to MET, affecting the jet transverse momentum $p_{\mathrm{T}}$ and the missing transverse momentum $p_{\mathrm{T}}^{\mbox{\footnotesize miss}}$. 
Meanwhile, the final state leptons are generally reconstructed before jets due to their clear signatures, so their $p_{\mathrm{T}}^\ell$ are rarely affected by the DPS effects. In addition, the final state jets will be affected little as soft hadrons are already included in the jet clustering. Note that the infrared safety of the jet algorithm~\cite{Salam:2007xv} guarantees that the clustered jets can be viewed as contributions from two separate parton-parton scatterings. 
Importantly, hadronic recoil is reconstructed primarily from jets. 
When performing hadronic calibrations using full samples of $Z\to\ell\ell$ events, the majority of the extracted information arises from final-state leptons. 
As mentioned above, such calibrations are driven almost entirely by lepton-related observables and rely primarily on their precise world-average values, such as the accurately known $Z$-boson mass. 
Consequently, the DPS effects are inherently excluded from the hadronic recoil calibration procedure. 

In the $M_W$ measurement, when tuning the MC simulation, the CDF collaboration calibrated the detector response to the visible hadronic recoil using only dilepton $Z$ events.
Specifically, a recoil response function is derived to map the hadronic recoil to the true transverse momentum $p_{\mathrm{T}}^{\ell\ell}$ of the $Z$ boson, thereby providing an approach to calibrate the hadronic recoil~\cite{CDF:2022hxs}. However, in high-DPS environments, the hadronic recoil response differs between the $W$- and $Z$-recoil templates, necessitating distinct calibration considerations for each mass measurement; see appendix~\ref{appendix:C}. For usual measurements, such as those of the $J/\psi$ meson mass, only a basic calibration of the lepton detector response is required, using leptons from $Z\to\ell\ell$ events. For $W$-boson measurements, however, further hadronic calibration is necessary. As shown in~\cite{CDF:2022hxs}, the CDF collaboration did not tune the modeling of hadronic activity using ``directly measured hadronic information'', which is sensitive to the DPS effects, in the recoil calibration procedure. In other words, relevant process-dependent DPS contributions will be partially lost in the $Z$-based calibration. Instead, they merely derived the recoil response utilizing the $u_\parallel$ data from $Z\to\ell\ell$ dilepton events, but without using the DPS-sensitive $u_\perp$ data, thereby obscuring the DPS contributions present in the $Z$-boson production; see appendix~\ref{appendix:C} for further details. Given the differences in realistic DPS modeling between the $W$- and $Z$-boson events, this $Z$-based calibration cannot fully account for all the DPS effects in $W$-boson production, which must be explicitly modeled and validated, with any residual contributions assigned as a systematic uncertainty; see the discussion in appendix~\ref{appendix:C}. Note that only by using the ``directly measured hadronic information'' can the calibrated recoil eliminate all the DPS contributions present in the $W$-boson production. Since the CDF analysis does not incorporate this information, its $M_W$ measurement inevitably retains the DPS contributions. To identify or mitigate DPS effects, it is essential to perform a calibration using direct hadronic measurements.

Given the high pile-up environment at the LHC, corrections for pile-up-induced effects are essential. 
The uniform pile-up environment enables jet calibration through correction factors~\cite{CMS:2011shu}, thereby mitigating these effects. 
Unlike other QCD radiations such as those from pile-up events, the production rate of the DPS radiations positively correlates with the cross section of the hard main or participant processes. Notably, the production rate of the pile-up radiations depends more on the intensity of beam collisions, relatively independent with the hard main processes. Although physically distinct, the DPS and pile-up effects similarly cause additional QCD radiation to the final state. 
Consequently, the DPS effects can be partly reduced by pile-up mitigation algorithms~\cite{CMS:2017yfk,Cacciari:2007fd,Soyez:2012hv,Butterworth:2008iy,Kogler:2018hem,Krohn:2009th,Bertolini:2014bba,Cacciari:2014gra,Dreyer:2018tjj}, which are designed to suppress pile-up effects through correction factors. Lacking a first-principles prediction, these correction factors are instead derived through data-driven procedures, involving some unknown degrees of freedom. For these calibration factors, the DPS and pile-up effects are almost indistinguishable, so the DPS effects are inadvertently corrected as a byproduct. As high-pile-up environments tend to have a smaller relative fraction of DPS activities compared to low-pile-up environments, the DPS effects after pile-up mitigation may differ significantly between experiments at different colliders, introducing subtle biases in the $M_W$ measurements and resulting in a discrepancy between the Tevatron and LHC. 

In principle, the $Z\to\ell\ell$ events can be utilized to enhance the sensitivity to DPS effects by excluding one of the two leptons from the reconstruction, thereby creating the $W$-like events. Thus far, this approach has been employed solely validate the techniques used in the $M_W$ measurement~\cite{CMS:2016nnd}, rather than to directly calibrate the $M_W$ determination. 
As in the analysis of all the $Z\to\ell\ell$ events, the impact of hadronic calibration on the $M_Z$ determination is too small to be of concern~\cite{LHCb:2025nob}, owing to its very low statistical weight. After excluding one lepton, the fractional hadronic contribution and low statistical weight remain similar, as they stem from the same underlying physics. As a sub-component of the already small hadronic contribution, the DPS-related QCD radiation should have an even smaller influence. Especially in high-pile-up environments, the DPS effects on the $W$-like $M_Z$ measurement will be further suppressed after the pile-up mitigation. 
Therefore, the $W$-like $M_Z$ measurement is not sensitive to the DPS effects. 
In addition, this approach suffers from reduced precision due to the exclusion of one lepton and the reduced event statistics. 
Hence, the uncertainty in the $W$-like measurement of $M_Z$ will be even larger than that in a standard $M_Z$ measurement. Currently, the lowest uncertainty is obtained with the $W^+$-like transverse mass, yielding a value of $\sim47$~MeV; see figure 6 of~\cite{CMS:2016nnd} for more details. In contrast, calibration demands a high degree of precision to ensure reliable measurements. Nevertheless, the $W$-like approach suffers from large uncertainties and low statistical weight, making it unsuitable for direct calibration of the $M_W$ value. 

Despite potentially larger statistical uncertainties, the $W$-like approach can offer an independent test for verifying the consistency of DPS effects across different processes, given its sensitivity to hadronic activity. Given that the uncertainties in the $M_W$ measurements at the Tevatron are now approaching those achieved at the LHC~\cite{CMS:2024lrd}, one could expect the uncertainty in a corresponding $W$-like $M_Z$ analysis using Tevatron data to be similarly around $\sim47$~MeV. 
As predicted in equation~\eqref{eq:DeltaM_Z}, the mass shift of $\Delta M_Z\sim85~\mbox{MeV}$ in the $W$-like $M_Z$ measurement, which is sensitive to the DPS threshold chosen, is already larger than the expected uncertainty, making it possible to verify the role of DPS effects across different high-precision measurements.

\section{Summary and Outlook}
\label{sec:5} 

In this article, we have demonstrated the DPS effects on the $W$-mass measurements and offered a compelling explanations for the observed $W$-boson anomaly. Specifically, using theoretical modeling and detector-level simulations, we show that the DPS effects can induce both a smearing and a hardening of the missing transverse momentum and transverse mass distributions. These changes produce an upward shift in the extracted $W$-boson mass, depending on DPS threshold $p_{\mathrm{T},\mbox{\footnotesize\rm cut}}^{\mbox{\scriptsize\rm DPS}}$. The resulting mass shift follows a power-law scaling, $\Delta M_W\sim60~\mbox{MeV}~(p_{\mathrm{T},\mbox{\footnotesize cut}}^{\mbox{\scriptsize DPS}}/10~\mbox{GeV})^{-3}$. For $p_{\mathrm{T},\mbox{\footnotesize\rm cut}}^{\mbox{\scriptsize\rm DPS}}\sim10$ GeV, this shift is $\sim60$ MeV, reconciling a significant portion of the observed discrepancy between the measured $W$-boson mass by CDF-II at the low-intensity Tevatron and the SM prediction. Interestingly, a similar mass shift can be expected in the $W$-like measurement of the $Z$-boson mass, a prediction that could be validated by further analysis of the Tevatron data. 

Note, the mass shift $\Delta M_W$, already larger than the QCD N$^3$LL+NNLO correction of $\sim10~\mathrm{MeV}$ to the $W$-boson mass, has long been overlooked in $M_W$ measurements, along with the DPS effects. 
At hadron colliders, these DPS effects are directly associated with the QCD activities at moderate $p_{\mathrm{T}}$. 
However, these semi-hard QCD activities may be suppressed in high-pile-up environments and relevant effects can be unintentionally mitigated, particularly after the pile-up corrections. This accounts for the $M_W$ discrepancy observed at the low-pile-up Tevatron, while no such tension is seen at the high-pile-up LHC. 

Additionally, the underlying semi-hard QCD activities cannot be identified directly. To reveal the DPS effects and distinguish them from the other factors such as pile-up and electronic noise, further analyses of the Tevatron data are necessary. These analyses should focus on addressing those semi-hard QCD effects, which may involve extending fitting intervals, varying the cutoffs in the jet algorithm, and improving the jet energy calibration. For instance, the PDS effects can be identified through precise $M_{W}$ and $M_{Z}$ measurements at different hadron colliders by varying jet transverse momentum cutoff $p_{\mathrm{T},\mbox{\footnotesize\rm cut}}^{j}$, thereby placing further constraints on DPS phenomenology. 

In hadron collisions, the DPS effects typically involve the presence of moderate-$p_{\mathrm{T}}$ jets. 
In general, jet measurements are less precise than leptonic ones at hadron colliders. 
It means that the DPS effects have a reduced influence on lepton-dominated measurements. 
To extract DPS effects, it is crucial to focus on measurements that are less dependent on leptonic signatures. 
For instance, the $W$-like events can be more sensitive to the DPS effects, as they involves $Z\to\ell\ell$ events with one lepton discarded to mimic the $W$ event topology. Further lepton-phobic measurements are expected offer crucial cross-validation in the semi-hard region, driving progress in our understanding of DPS dynamics. 
    
Likewise, other measurements, such as those of the Drell-Yan process, are sensitive to the DPS effects. 
Specifically, due to the DPS effects, the total cross section for any inclusive process can be enhanced by $\sim10^{-2}$, and the distribution of missing transverse momenta can be hardened by $\mathcal{O}(10^{-2})$ to $\mathcal{O}(10^{-1})~\mbox{GeV}$. Upcoming measurements for rare processes at the LHC, including vector boson scattering, di-Higgs production, $t\overline{t}V$ production, flavor changing neutral current processes involving top quarks, are all potentially affected by the cross-section enhancements. 
Also, other precision measurements, such as the top quark mass, spin correlation and resonance shape in the $t\overline{t}$ production process, as well as the $W$-like measurement of the $Z$-boson mass proposed above, are highly sensitive to the performance of missing transverse momentum reconstruction, making them potentially vulnerable to the DPS effects. In the future, the DPS effects are expected to be probed through more high-precision measurements. 

So far our study has been presented based on the factorized cross-section formula, a standard approach common to both theoretical and experimental studies. 
It should be noted that this formula can be derived from some simplified DPS models by neglecting longitudinal momentum correlations between the two partons and imposing the constraint that their combined momentum does not exceed that of the parent proton~\cite{Cao:2017bcb}. 
Such a study serves to assess the applicability of these simplified DPS models in the semi-hard regime, examine the consistency of the model parameters, such as the DPS threshold and effective cross-section, across different experiments, and provide an experimental foundation for investigating key theoretical assumptions, such as factorization. For a more visual perspective, we use the standard factorized cross-section formula to demonstrate how the DPS effects influence precise $M_{W}$ measurements and vary as a function of the DPS threshold. Conversely, high-precision $M_W$ measurements are expected to provide a powerful probe for testing and constraining phenomenological DPS models in the near future. 

Finally, it is worth mentioning that the $W$-boson mass can also be shifted by the presence of new particles or anomalous electroweak interactions. However, such new physics will change the measured $W$-boson mass consistently across different experiments. Consequently, discrepancies between measurements likely stem from differences in their respective experimental environments. At present, the DPS effects provide a compelling explanation for the disagreement between the Tevatron and LHC measurements.

\section*{Acknowledgements}
We thank Qing-Hong Cao, Qiang Li, Yandong Liu, Meng Lu, Wei-Min Song, Bin Yan, Zhao-Huan Yu, Hao Zhang for useful discussions. We also thank the anonymous reviewer for their constructive comments. This work is supported by the National Key R\&D Program of China (Grant Nos. 2025YFF0521800, 2025YFF0511100, and 2021YFA0718500), the Chinese Academy of Sciences (Grant No. E32983U810), the National Natural Science Foundation of China (Grant No. 12235001), and China's Space Origins Exploration Program.

\section*{Data Availability} 
All the data are available in the paper. 
\section*{Competing Interests} 
The authors declare that they have no conflicts of interest to this work. \\
\\

\bibliography{draft}

\begin{thebibliography}{10}
\bibitem{Herget:2018uam}
V.~Herget, \emph{{A novel approach for the calibration of the hadronic recoil
  for the measurement of the mass of the W boson with the ATLAS Experiment}},
  Ph.D. thesis, U. Wurzburg (main), 2018.
\end{thebibliography}
\bibliographystyle{JHEP}


\appendix

\newpage

\section{Formulating DPS effects on $M_W$ measurements}
\label{appendix:A}\vspace*{-0.3cm}

Here we present the calculation details for the $p_{\mathrm{T}}^{\mbox{\scriptsize miss}}$ and $m_{\mathrm{T}}$ distributions of the signal events from the DPS-only processes. 
In realistic measurements, as the distributions of $p_{\mathrm{T}}^\ell$, $p_{\mathrm{T}}^{\mbox{\scriptsize miss}}$ and $m_{\mathrm{T}}$ do not show much difference for the electron and muon channels, 
we do not need to distinguish these two channels. So we simply combine these two channels and obtain the normalized distributions. 
We extract the $p_{\mathrm{T}}^\ell$,  $p_{\mathrm{T}}^{\mbox{\scriptsize miss}}$ and $m_{\mathrm{T}}$ distributions from the $W$ boson production in the CDF-II data~\cite{CDF:2022hxs}. Then, we can fit these normalized distributions with simplified Landau distribution~\cite{Landau:1944fvs}
\begin{align}\label{eq:fit}
Z\equiv Z(p; N,x,a,b,c)
=\frac{1}{N}\exp\left[-\frac{(p-xM_W)^2}{
\big(a+b(p-xM_W)+c(p-xM_W)^2\big)^2}\right]\,, 
\end{align}
where $p=p_{\mathrm{T}}^\ell,~p_{\mathrm{T}}^{\mbox{\scriptsize miss}},~m_{\mathrm{T}}$ for their respective distributions, as well as $N,\,x,\,a,\,b$, and $c$ are five free fitting parameters. 
Generally, one has
\begin{align}
\frac{\mathrm{d}\sigma^{\mbox{\scriptsize{SPS}}}}{\sigma^{\mbox{\scriptsize{SPS}}}\mathrm{d} p_{\mathrm{T}}^\ell}\equiv A(p_{\mathrm{T}}^\ell)\,,\,
\frac{\mathrm{d}\sigma^{\mbox{\scriptsize{SPS}}}}{\sigma^{\mbox{\scriptsize{SPS}}}\mathrm{d} p_{\mathrm{T}}^{\mbox{\scriptsize miss}}}\equiv B(p_{\mathrm{T}}^{\mbox{\scriptsize miss}})\,,\,
\frac{\mathrm{d}\sigma^{\mbox{\scriptsize{SPS}}}}{\sigma^{\mbox{\scriptsize{SPS}}}\mathrm{d} m_{\mathrm{T}}}\equiv C(m_{\mathrm{T}})\,, 
\end{align}
where $A$, $B$, and $C$ can be well described by the functional form $Z=Z(p; N,x,a,b,c)$ in the data fitting. 
These equations illustrate how these normalized distributions relate to the cross sections for their respective physical processes. 
From the fits, we obtain the best-fit values of the parameters, as shown in table~\ref{tab:fit}. 
Here, the $W$ mass is chosen to align with the SM prediction, i.e., $M_W=M_W^{\mbox{\scriptsize SM}}=80359.1~\mbox{MeV}$\footnote{Here, for any distribution $Z$, we first obtain $s\equiv xM_W^{\mbox{\scriptsize SM}}$ from data fits. 
Then in the event analysis, we obtain the best-fit values of $t\equiv x(M_W^{\mbox{\scriptsize SM}}+\Delta M_W)$ by fitting the $p_{\mathrm{T}}^{\mbox{\scriptsize miss}}$ and $m_{\mathrm{T}}$ distributions of both SPS and DPS events. Finally, we compute the $W$-boson mass shift, i.e., $\Delta M_W=(t/s-1)M_W^{\mbox{\scriptsize SM}}$, which is insensitive to the $M_W^{\mbox{\scriptsize SM}}$ value chosen. }. 
Theoretically, we can adjust the $M_W$ value in the functions $Z=A,\,B,\,C$ so as to obtain their distributions for any $M_W$, which can be used as SPS templates to the fit the distributions combined with both SPS and DPS events. 
These $Z$ distributions are consistent with those extracted from the best SPS simulations by the CDF collaboration within the experimental uncertainties, where the simulations are calculated from first principles under the SPS-only assumption. 
Hence, they can serve as templates for describing the SPS processes, also referred as the SPS templates in the main text.

\begin{table}[ht]
\centerline{
\begin{tabular}{c|c|c|c|c|c}
$Z$&$N$&$x$&$a$&$b$&$c$\\
\hline
~~$A$~~&~~11.08~~&~~0.4661~~&~~6.418~~&~~-0.1118~~&~~0.01043~\\
\hline
~~$B$~~&~~12.34~~&~~0.4603~~&~~7.880~~&~~-0.06364~~&~~0.002705~\\
\hline
~~$C$~~&~~19.10~~&~~0.9145~~&~~11.15~~&~~0.1256~~&~~-0.003254~\\
\hline
~~$D$~~&~~8.839~~&~~$0.0524$~~&~~4.362~~&~~0.3924~~&~~-0.007343~\\
\end{tabular}
}
\caption{
Best-fit parameters from the experimental results and the simulated dijet events. 
Each line shows a set of parameters for the distribution in the form of equation~\eqref{eq:fit}. 
Note that $N$, $x$, $a$, $b$, and $c$ are in units of GeV, 1, GeV, 1, $\mbox{GeV}^{-1}$, respectively. 
For $X=D$, we choose a cut of $p_{\mathrm{T}}^j>10~\mbox{GeV}$ for the generated events in the dejet process for example. }
\label{tab:fit}
\end{table}

We obtain the $p_{\mathrm{T},jj}^{\mbox{\scriptsize miss}}$ distribution of spectator events from the dijet production simulations. 
This distribution
\begin{align}\label{eq:dijet}
\frac{\mathrm{d}\sigma_{jj}}{\sigma_{jj}\mathrm{d} p_{\mathrm{T},jj}^{\mbox{\scriptsize miss}}}= D(p_{\mathrm{T},jj}^{\mbox{\scriptsize miss}})\,, 
\end{align}
can also be well fitted in the same form as equation~\eqref{eq:fit}, where $p=p_{\mathrm{T},jj}^{\mbox{\scriptsize miss}}$. 
For example, we apply a cut of $p_{\mathrm{T}}^j>10~\mbox{GeV}$ on jets at the generator level. 
In this specific case, we fit the $p_{\mathrm{T},jj}^{\mbox{\scriptsize miss}}$ distribution with the function $D$, and obtain a set of the best-fit parameters; see table~\ref{tab:fit} for details. 
Note that $x=0.0524$ corresponds to $xM_W=4.215~\mbox{GeV}$, which is where the peak is located in the $p_{\mathrm{T},jj}^{\mbox{\scriptsize miss}}$ distribution. 
Note that the value $\langle p_{\mathrm{T},jj}^{\mbox{\scriptsize miss}}\rangle\sim4.2~\mbox{GeV}$ is from the Delphes fast simulation. 
Delphes was tuned to provide a good approximation for the ATLAS and CMS detectors, but it has never been tuned for the CDF detector. 
So it is worth examining whether our simulation can accurately model the CDF detector using the CDF-II data. 
Generally, the $p_{\mathrm{T}}^{\mbox{\scriptsize miss}}$ reconstruction involves using tracks to recover momenta from low-$p_{\mathrm{T}}$ charged particles that may not be detected by the calorimeters. 
Therefore, the $p_{\mathrm{T}}^{\mbox{\scriptsize miss}}$ distribution from low-$p_{\mathrm{T}}$ jets is primarily influenced by calorimeter reactions rather than by specific dijet or diphoton processes. 
As shown by the CDF-II data for the diphoton process, the $p_{\mathrm{T}}^{\mbox{\scriptsize miss}}$ distribution peaks at $\sim4.9~\mbox{GeV}$~\cite{Merritt:1996wy}, which is consistent with our simulation. 
Therefore, our simulation can be used to describe the $p_{\mathrm{T}}^{\mbox{\scriptsize miss}}$ distribution from the CDF dijet processes. 
If $\langle p_{\mathrm{T},jj}^{\mbox{\scriptsize miss}}\rangle$ increases to $\sim4.9~\mbox{GeV}$ from $\sim4.2~\mbox{GeV}$, the smearing resolutions of $p_{\mathrm{T}}^{\mbox{\scriptsize miss}}$ and $m_{\mathrm{T}}$ will increase by $\sim17\%$, and consequently, the influence of the DPS effects will become more significant.

Let us turn back to signal $W$ production at the Tevatron. 
The DPS process can be viewed as concurrence of an SPS $W$ production process, designated as participant process $P$, and an SPS dijet process, termed as spectator process $S$. 
Now we can extract the distributions of $Z=A,\,B,\,C$ from process $P$ by fitting the experimental data, and obtain the distribution $Z=D$ from the simulated events generated in the process $S$ with {\tt MadGraph5}. 
Thus, we could get the DPS distributions by combining the events from the two SPS processes $P$ and $S$. 
Clearly, the missing transverse momentum $p_{\mathrm{T},\mbox{\scriptsize DPS}}^{\mbox{\scriptsize miss}}$ from a DPS event can be expressed as
\begin{align}
\label{eq:p_DPS^miss}
p_{\mathrm{T},\mbox{\scriptsize DPS}}^{\mbox{\scriptsize miss}}=\sqrt{(p_{\mathrm{T},\mbox{\scriptsize SPS}}^{\mbox{\scriptsize miss}})^2+(p_{\mathrm{T},jj}^{\mbox{\scriptsize miss} })^2+2p_{\mathrm{T},\mbox{\scriptsize SPS}}^{\mbox{\scriptsize miss}}p_{\mathrm{T},jj}^{\mbox{\scriptsize miss}}\cos\alpha}\,, 
\end{align}
where $\alpha$ is the open angle between $\vec{p}_{\mathrm{T},\mbox{\scriptsize SPS}}^{~\mbox{\scriptsize miss}}$ and $\vec{p}_{\mathrm{T},jj}^{~\mbox{\scriptsize miss}}$. 
Then the $p_{\mathrm{T},\mbox{\scriptsize DPS}}^{\mbox{\scriptsize miss}}$ distribution reads
\begin{align}
\label{eq:DPSpmissD}
\frac{\mathrm{d}\sigma^{\mbox{\scriptsize{DPS}}}}{\sigma^{\mbox{\scriptsize{DPS}}}\mathrm{d} p_{\mathrm{T},\mbox{\scriptsize DPS}}^{\mbox{\scriptsize miss}}}
=&\frac{1}{\pi}\int B(p_{\mathrm{T},\mbox{\scriptsize SPS}}^{\mbox{\scriptsize miss}})\mathrm{d} p_{\mathrm{T},\mbox{\scriptsize SPS}}^{\mbox{\scriptsize miss}}D(p_{\mathrm{T},jj}^{\mbox{\scriptsize miss}})\mathrm{d} p_{\mathrm{T},jj}^{\mbox{\scriptsize miss}}\nonumber\\
\vspace*{-5mm}
&\times\frac{p_{\mathrm{T},\mbox{\scriptsize DPS}}^{\mbox{\scriptsize miss}}}{p_{\mathrm{T},\mbox{\scriptsize SPS}}^{\mbox{\scriptsize miss}}p_{\mathrm{T},jj}^{\mbox{\scriptsize miss}}}\left\{1-\left[\frac{(p_{\mathrm{T},\mbox{\scriptsize DPS}}^{\mbox{\scriptsize miss}})^2-(p_{\mathrm{T},\mbox{\scriptsize SPS}}^{\mbox{\scriptsize miss} })^2-p_{\mathrm{T},jj}^{\mbox{\scriptsize miss} 2}}{2p_{\mathrm{T},\mbox{\scriptsize SPS}}^{\mbox{\scriptsize miss}}p_{\mathrm{T},jj}^{\mbox{\scriptsize miss}}}\right]^2\right\}^{-1/2}\,. 
\end{align}
Similarly, the transverse mass $m_{\mathrm{T},\mbox{\scriptsize DPS}}$ of a DPS event is: 
\begin{align}
\label{eq:DPSmt}
(m_{\mathrm{T},\mbox{\scriptsize DPS}})^2=(m_{\mathrm{T},\mbox{\scriptsize SPS}})^2+2p_{\mathrm{T},\mbox{\scriptsize SPS}}^\ell (p_{\mathrm{T},\mbox{\scriptsize DPS}}^{\mbox{\scriptsize miss}}-p_{\mathrm{T},\mbox{\scriptsize SPS}}^{\mbox{\scriptsize miss}})
-2p_{\mathrm{T},\mbox{\scriptsize SPS}}^\ell\, p_{\mathrm{T},jj}^{\mbox{\scriptsize miss}}\cos\beta\,, 
\end{align}
where $\beta$ is the open angle between $\vec{p}_{\mathrm{T},\mbox{\scriptsize SPS}}^{~\ell}$ and $\vec{p}_{\mathrm{T},jj}^{~\mbox{\scriptsize miss}}$. 
As the spectator processes, such as the dijet production process, are unable to alter the measured transverse momentum $p_{\mathrm{T}}^\ell$ of electrons or muons, $p_{\mathrm{T},\mbox{\scriptsize SPS}}^\ell$ can be treated the same as $p_{\mathrm{T}}^\ell$. 
Note, $p_{\mathrm{T},\mbox{\scriptsize DPS}}^{~\mbox{\scriptsize miss}}$ depends on the angle $\alpha$, which is highly correlated with $\beta$. 
In the laboratory frame, $\vec p_{\mathrm{T},\mbox{\scriptsize SPS}}^{~\ell}$ and $\vec p_{\mathrm{T},\mbox{\scriptsize SPS}}^{~\mbox{\scriptsize miss}}$ are approximately back to back. 
Thus, we fix $\alpha=\pi-\beta$ for simplicity. 
Additionally, we estimate the joint distributions of $p_{\mathrm{T},\mbox{\scriptsize SPS}}^\ell$ and $p_{\mathrm{T},\mbox{\scriptsize SPS}}^{\mbox{\scriptsize miss}}$ from the SPS processes as the product of two independent distributions $A$ and $B$. 
Accordingly, the $m_{\mathrm{T},\mbox{\scriptsize DPS}}$ distributions are calculated as
\begin{align}
\label{eq:DPSmtD}
\frac{\mathrm{d}\sigma^{\mbox{\scriptsize{DPS}}}}{\sigma^{\mbox{\scriptsize{DPS}}}\mathrm{d} m_{\mathrm{T},\mbox{\scriptsize DPS}}}
=&\int A(p_{\mathrm{T}}^\ell)\mathrm{d} p_{\mathrm{T}}^\ell B(p_{\mathrm{T},\mbox{\scriptsize SPS}}^{\mbox{\scriptsize miss}})\mathrm{d} p_{\mathrm{T},\mbox{\scriptsize SPS}}^{\mbox{\scriptsize miss}} D(p_{\mathrm{T},jj}^{\mbox{\scriptsize miss}})\mathrm{d} p_{\mathrm{T},jj}^{\mbox{\scriptsize miss}}\nonumber\\
&\times\int_0^\pi\frac{\mathrm{d}\beta}{\pi}C(m_{\mathrm{T},\mbox{\scriptsize SPS}})\frac{m_{\mathrm{T},\mbox{\scriptsize DPS}}}{m_{\mathrm{T},\mbox{\scriptsize SPS}}}\Bigg{|}_{m_{\mathrm{T},\mbox{\scriptsize SPS}}=M_{\mathrm{T}}}, 
\end{align}
where $M_{\mathrm{T}}$ is the solution to $m_{\mathrm{T},\mbox{\scriptsize SPS}}$ of equation~\eqref{eq:DPSmt}: 
\begin{align*}
M_{\mathrm{T}}^2=&(m_{\mathrm{T},\mbox{\scriptsize DPS}})^2+2p_{\mathrm{T}}^\ell p_{\mathrm{T},jj}^{\mbox{\scriptsize miss}}\cos\beta+2p_{\mathrm{T}}^\ell p_{\mathrm{T},\mbox{\scriptsize SPS}}^{\mbox{\scriptsize miss}}\nonumber\\
&-2p_{\mathrm{T}}^\ell\sqrt{(p_{\mathrm{T},\mbox{\scriptsize SPS}}^{\mbox{\scriptsize miss}})^2+(p_{\mathrm{T},jj}^{\mbox{\scriptsize miss}})^2-2p_{\mathrm{T},\mbox{\scriptsize SPS}}^{\mbox{\scriptsize miss}}p_{\mathrm{T},jj}^{\mbox{\scriptsize miss}}\cos\beta}\,. 
\end{align*}

Following the CDF collaboration, we fit the SPS distributions in the region
\begin{align*}
&30~\mbox{GeV}<p_{\mathrm{T}}^\ell,p_{\mathrm{T}}^{\mbox{\scriptsize miss}}<55~\mbox{GeV}\,,\\
&60~\mbox{GeV}<m_{\mathrm{T}}<100~\mbox{GeV}\,. 
\end{align*}
In the calculation, we slightly extend the integral intervals to
\begin{align*}
&20~\mbox{GeV}<p_{\mathrm{T}}^\ell,p_{\mathrm{T}}^{\mbox{\scriptsize miss}}<60~\mbox{GeV}\,,\\
&50~\mbox{GeV}<m_{\mathrm{T}}<110~\mbox{GeV}\,, 
\end{align*}
in order to smooth the DPS distributions at their two ends of the fitting intervals. 

As already demonstrated in the main text, the DPS effects can alter the $p_{\mathrm T,\mbox{\scriptsize DPS}}^{\mbox{\scriptsize miss}}$ distribution, thus leading to the changes in the distributions of the measured $p_{\mathrm T}^{\mbox{\scriptsize miss}}$ and $m_{\mathrm T}$, as indicated by Eqs.~\eqref{eq:ptmiss} and \eqref{eq:mT_DPS}. 
Note that these changes primarily stem from the smearing, and this DPS effect can alter the fitting results. 
When we use a SPS template to fit some distribution, the DPS smearing would distort the distribution and change the fitting results. 
In fact, due to a lack of fully understanding, the quantitative description of the DPS effects has larger uncertainties than those present in the SPS template. 
Thus, we have to use the SPS template to fit various distributions and measure the $W$-boson mass. This will induce an apparent $M_W$ shift to accommodate the ignored DPS effects. 

On the other hand, the DPS effects may affect the production processes of $Z$ bosons as well as $J/\psi$ and $\Upsilon$ particles, which are used to calibrate the momenta and energies of the final-state particles like the leptons~\cite{CDF:2022hxs}. 
However, contrast to the $W$ boson production, all the final-state particles used in the calibrations are visible. Note that the DPS effects mainly play a role in affecting the detections and measurements of missing transverse energies. 
Since no missing transverse energies are involved, the calibrations to the observables like $p_{\mathrm{T}}^\ell$ cannot be changed by the DPS effects. 
Thus, the CDF calibrations do not help to identify the DPS effects.

However, if there is bilateral symmetry in the distribution within the fitting interval selected, the smearing effect on these fitting results will be rather small. For example, if we choose another fitting interval such as $32~\mbox{GeV}<p_{\mathrm T}^{\mbox{\scriptsize miss}}<42~\mbox{GeV}$, the $W$-boson mass shift induced by the DPS effects could be reduced. It may be beneficial to only fit distributions over intervals where the distributions are obtained by cutting their high-energy tails off and they become bilaterally symmetric with respect to their peaks, which is far beyond the scope of this work. However, this may lead to a larger uncertainty in the fitted $M_W$, and it does not aid in understanding the physics behind the tension observed in the CDF-II $M_W$ measurement with the SM prediction.

\section{Measuring the $Z$-boson mass in $W$-like events}
\label{appendix:B}\vspace*{-0.3cm}

As mentioned above, the $W$-boson mass is measured using a sample of $W\to\ell\nu$ events, in which the neutrino $\nu$ escapes detection, resulting in missing transverse momentum $p_{\mathrm{T}}^{\mbox{\scriptsize miss}}$. 
Similarly, the $Z$-boson mass $M_Z$ can be determined from the so-called $W$-like events that mimic the $W\to\ell\nu$ topology, namely $Z\to\ell\ell$ decays in which one of the two charged leptons is not reconstructed~\cite{CMS:2016nnd}, leading to a single visible lepton and apparent missing transverse momentum. 
In fact, the $W$-like event sample can be derived from the $Z\to\ell\ell$ background in the $M_W$ measurements, with one of the two leptons not reconstructed. 
This yields a sample of 304,714 events. 
Accordingly, we extract the $p_{\mathrm{T}}^\ell$,  $p_{\mathrm{T}}^{\mbox{\scriptsize miss}}$ and $m_{\mathrm{T}}$ distributions for the $W$-like events from the CDF-II data~\cite{CDF:2022hxs}. 
Then, we can fit their normalized distributions with
\begin{align}\label{eq:fitZ}
Z^\prime\equiv Z^\prime(p; N^\prime,y^\prime,a^\prime,b^\prime,c^\prime)
=\frac{1}{N^\prime}\exp\left[-\frac{(p-y^\prime M_Z)^2}{
\big(a^\prime+b^\prime(p-y^\prime M_Z)+c^\prime(p-y^\prime M_Z)^2\big)^2}\right]\,, 
\end{align}
where $p$ may be $p_{\mathrm{T}}^\ell,~p_{\mathrm{T}}^{\mbox{\scriptsize miss}}$ or $m_{\mathrm{T}}$, as well as $N,\,y^\prime,\,a^\prime,\,b^\prime$, and $c^\prime$ are five free fitting parameters. Note that it takes the same form as equation~\eqref{eq:fit}. 
For the $Z$ boson production, 
\begin{align}
\frac{\mathrm{d}\sigma^{\mbox{\scriptsize{SPS}}}}{\sigma^{\mbox{\scriptsize{SPS}}}\mathrm{d} p_{\mathrm{T}}^\ell}\equiv A^\prime(p_{\mathrm{T}}^\ell)\,,\,
\frac{\mathrm{d}\sigma^{\mbox{\scriptsize{SPS}}}}{\sigma^{\mbox{\scriptsize{SPS}}}\mathrm{d} p_{\mathrm{T}}^{\mbox{\scriptsize miss}}}\equiv B^\prime(p_{\mathrm{T}}^{\mbox{\scriptsize miss}})\,,\,
\frac{\mathrm{d}\sigma^{\mbox{\scriptsize{SPS}}}}{\sigma^{\mbox{\scriptsize{SPS}}}\mathrm{d} m_{\mathrm{T}}}\equiv C^\prime(m_{\mathrm{T}})\,, 
\end{align}
where $A^\prime$, $B^\prime$, and $C^\prime$ take the form $Z^\prime= Z^\prime(p; N^\prime,y^\prime,a^\prime,b^\prime,c^\prime)$. 
According to the SM prediction, the $Z$-boson mass is set to be $M_Z^{\mbox{\scriptsize{SM}}}=91204.7~\mbox{MeV}$, as obtained from the global fit to precision electroweak data~\cite{deBlas:2021wap}. We now extract the distributions $A^\prime$, $B^\prime$, and $C^\prime$ from process $P$ by fitting the experimental data, and obtain distribution $D$ from the SPS dijet process $S$ using {\tt MadGraph5} simulations. 
Thus, by combining the events from the two SPS processes $P$ and $S$, we calculate the distributions of $p_{\mathrm{T},\mbox{\scriptsize DPS}}^{\mbox{\scriptsize miss}}$~\eqref{eq:DPSpmissD} and $m_{\mathrm{T},\mbox{\scriptsize DPS}}$~\eqref{eq:DPSmtD} for the DPS-only processes. Then, we derive the total $p_{\mathrm{T}}^\ell,~p_{\mathrm{T}}^{\mbox{\scriptsize miss}}$ and $m_{\mathrm{T}}$ distributions of the $W$-like events, including both SPS-only and DPS-only events; see table~\ref{tab:fitZ} for further details. Consequently, for a given $p_{\mathrm{T},\mbox{\footnotesize cut}}^{\mbox{\footnotesize DPS}}$, we fit each total distribution using the function in equation~\eqref{eq:fitZ} to extract a $Z$-boson mass value, and then form a weighted average by assuming the weights are the same as the $M_W$ measurements. Finally, analogously to the $W$-boson mass, we also observe a mass shift in the $Z$-boson mass: 
\begin{align}
\label{eq:DeltaM_Z}
\Delta M_Z=\Delta_{10}^\prime~\,(p_{\mathrm{T},\mbox{\footnotesize cut}}^{\mbox{\footnotesize DPS}}/10~\mbox{GeV})^{-3}\,, 
\end{align}
where the value of $\Delta_{10}^\prime$ is $82^{+14}_{-14}~\mbox{MeV}$ in the CMS analysis, with $\sigma_{\mbox{\footnotesize eff}}=12.2^{+2.9}_{-2.2}~\mbox{mb}$~\cite{CMS:2022pio}, and $92^{+15}_{-11}~\mbox{MeV}$ in the ATLAS analysis, with $\sigma_{\mbox{\footnotesize eff}}=10.6\pm1.8~\mbox{mb}$~\cite{ATLAS:2025bcb}. 
Note that it decreases with the DPS threshold $p_{\mathrm{T},\mbox{\footnotesize cut}}^{\mbox{\footnotesize DPS}}$ in the same power-law pattern as that observed for the $W$-boson mass.

\begin{table}[ht]
\centerline{
\begin{tabular}{c|c|c|c|c|c}
$Z^\prime$&$N^\prime$&$y^\prime$&$a^\prime$&$b^\prime$&$c^\prime$\\
\hline
~~$A^\prime$~~&~~14.63~~&~~0.4122~~&~~10.68~~&~~-0.2957~~&~~0.01009~\\
\hline
~~$B^\prime$~~&~~12.91~~&~~0.3736~~&~~9.596~~&~~-0.3042~~&~~-0.01915~\\
\hline
~~$C^\prime$~~&~~23.12~~&~~0.7676~~&~~13.79~~&~~0.5585~~&~~-0.01463~\\
\end{tabular}
}
\caption{
Same as in table~\ref{tab:fit}, but for the $W$-like events. 
}
\label{tab:fitZ}
\end{table}

\begin{figure}
\centerline{
\includegraphics[width=\columnwidth]{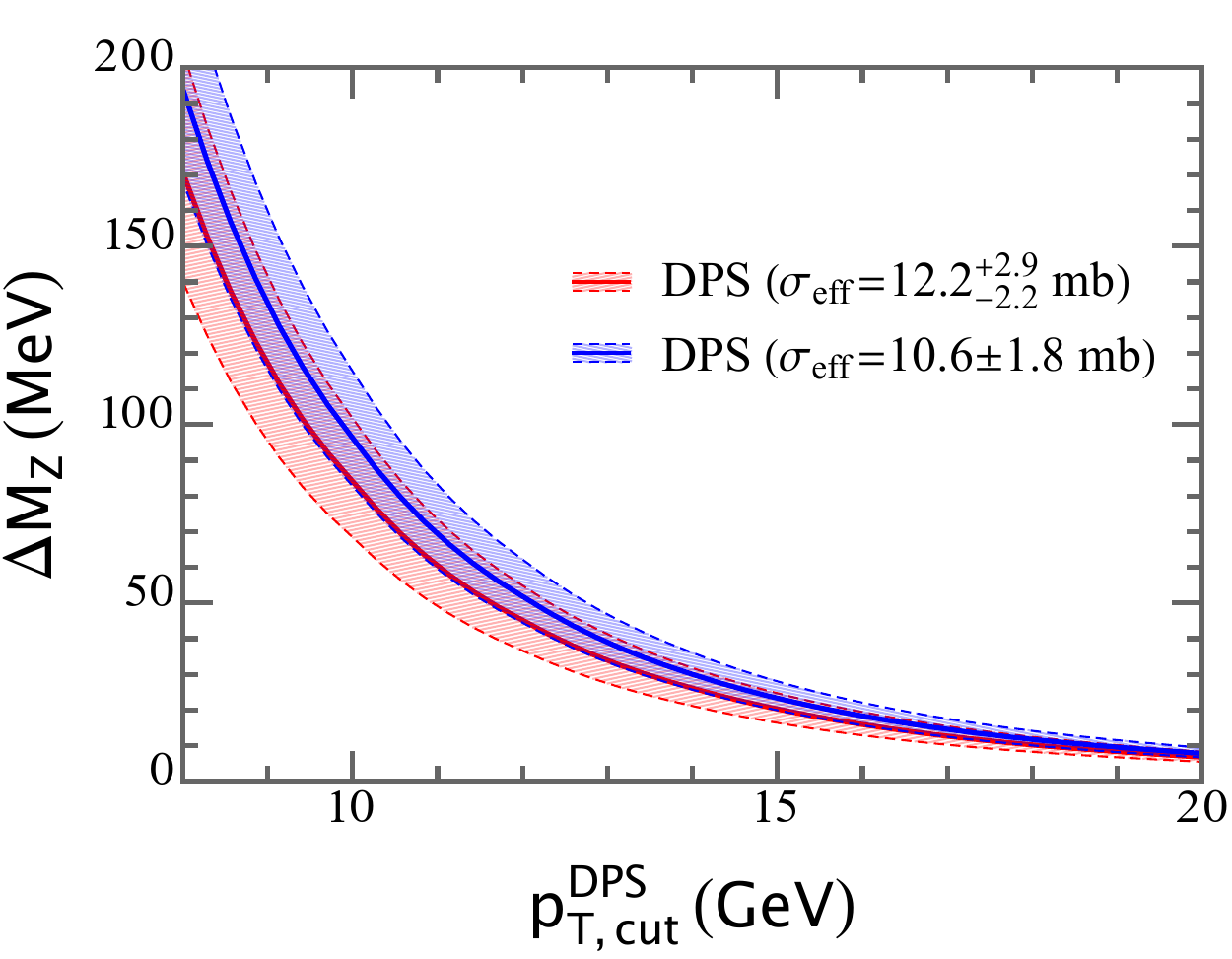}
}
\caption{The mass shift $\Delta M_Z$ as a function of the DPS threshold $p_{\mathrm{T},\mbox{\footnotesize cut}}^{\mbox{\footnotesize DPS}}$, obtained from the fits to the $W$-like events, with all other settings and conventions as in figure~\ref{fig:2}. 
}
\label{fig:3}
\end{figure}

The $W$-like measurement is blind to dilepton kinematics (e.g., invariant mass $m_{\ell\ell}$, transverse momentum $p_{\mathrm{T}}^{\ell\ell}$; see section~\ref{sec:4}), making the results particularly sensitive to missing transverse momentum $p_{\mathrm{T}}^{\mbox{\footnotesize miss}}$ associated with the removed lepton. 
Thus, it provides a unique probe of DPS effects, as these arise from hadronic interactions rather than dilepton kinematics. 
Compared to the high-luminosity LHC, the DPS effects are more easily observable at the Tevatron owing to its low-pile-up environment; see section~\ref{sec:4} for details. 
Furthermore, hadronic recoil may also contribute to the uncertainty in missing transverse momentum. 
Given that hadronic recoil information from $Z$ boson events is not fully utilized as an input in standard calibration procedures (e.g., the CDF-II $M_{W}$ measurement~\cite{CDF:2022hxs}), potential differences in hadronic recoil between $W$ and $Z$ events have not been thoroughly investigated at the Tevatron. 
Therefore, the implementation of the $W$-like measurements aids in verifying the role of DPS effects, potentially serving as a key step forward in addressing the $W$-boson mass anomaly.

\section{Supplementrary Comments on Recoil Calibration}
\label{appendix:C}\vspace*{-0.3cm}

In particle physics experiments, particularly at hadron colliders such as the Tevatron and the LHC, the $p_{\mathrm{T}}$-balance is used for calibrating detector responses, verifying energy/momentum conservation, and assessing systematic uncertainties. 

In hadron-hadron collisions, the net initial transverse momentum $\vec{p}_{\mathrm{T}}^{\text{initial}}$ is zero. Consequently, the vector sum of the transverse momenta of all final-state particles must also be zero: 
\begin{align}
\vec{p}_{\mathrm{T}}^{~\mbox{\scriptsize initial}}=0 \implies \sum\vec{p}_{\mathrm{T}}^{~\mbox{\scriptsize final}}=0\,. 
\end{align}
If a process produces a high-$p_{\mathrm{T}}$ object, such as a $Z$ boson, a photon, or a $W$ boson, there must be a recoil momentum that balances it in the opposite direction.

In the $Z\to\ell\ell$ events, the momenta of the two leptons are measured with high precision. 
In general, the true transverse hadronic recoil momentum, $\vec{u}_{\mathrm{T}}^{~\mbox{\scriptsize true}}$, is generally defined by momentum conservation: 
\begin{align}
\label{eq:balance_CDF}
\vec{u}_{\mathrm{T}}^{~\mbox{\scriptsize true}}=-(\vec{p}_{\mathrm{T}}^{~\ell_1}+\vec{p}_{\mathrm{T}}^{~\ell_2/\nu_2})=-\vec{p}_{\mathrm{T}}^{~V}\,, 
\end{align}
where $V$ denotes either the $Z$ or $W$ boson. For $V=Z$, it serves as the basis for deriving the recoil response function $R\equiv\vec{u}_{\mathrm{T}}\cdot\hat{u}_{\mathrm{T}}^{\mbox{\scriptsize true}}/u_{\mathrm{T}}^{\mbox{\scriptsize true}}$ in the CDF $Z\to\ell\ell$ analysis~\cite{CDF:2022hxs}. Here, $\vec{u}_{\mathrm{T}}$ denotes the hadronic recoil and $\hat{u}_{\mathrm{T}}$ the corresponding unit vector. 
If a $W$ boson were produced without extra spectator radiation, $\vec{p}_{\mathrm{T}}^{~W}$ and $\vec{u}_{\mathrm{T}}$ would be in perfect balance. Under this idealized condition, the $Z$-derived $R$ can be directly applied to $M_{W}$ measurements.

In the 2022 $M_{W}$ measurement~\cite{CDF:2022hxs}, the CDF collaboration applied a data-driven recoil response $R$ derived from $Z\to\ell\ell$ events to tune the $W\to\ell\nu$ MC simulation to data~\cite{CDF:2007mxw}. Generally, the hadronic recoil $\vec{u}_{\mathrm{T}}$ is decomposed into parallel ($u_\parallel$) and perpendicular ($u_\perp$) components along the direction of $\vec{p}_{\mathrm{T}}^{~Z}$ (i.e., $\vec{p}_{\mathrm{T}}^{~\ell\ell}$). In the current CDF analysis, the recoil response function $R$ is parametrized as a function of the $Z$ boson $p_{\mathrm{T}}^{\ell\ell}$. For $Z\to\ell\ell$ events, they rescales the measured $u_\parallel$ to  match the measured $\vec{p}_{\mathrm{T}}^{~\ell\ell}$ using $Z$-calibrated response function $R$. Similarly, for $W\to\ell\nu$ events, the unknown hadronic recoil $\vec{u}_{\mathrm{T}}$ is modeled by using the same response function $R$ to the generator-level $\vec{p}_{\mathrm{T}}^{~W}$ from the MC simulation and generating MC samples of $W\to\ell\nu$ events to emulate the $W$-boson data. Note that in the CDF recoil calibration, only $\vec{p}_{\mathrm{T}}^{~\ell\ell}$ is replaced with the theoretically derived $\vec{p}_{\mathrm{T}}^{~W}$, and $R$ is extracted exclusively from the $u_\parallel$ data. 

If there were no significant differences between the $W$- and $Z$-recoil templates, the $Z$-calibrated $R$ could partly integrate the DPS contributions in $W$-boson production, even without a priori understanding of the hadronic recoil. Since the magnitudes of $u_{\parallel}$ and $u_{\perp}$ are both at the same order as $p_{\mathrm{T}}^Z\sim4$ GeV~\cite{CDF:2022hxs}, both components should be carefully accounted for when modeling the DPS effects in the $Z$-boson data. Compared to $u_{\perp}$, $u_{\parallel}$ is easily affected by the lepton measurements and become insensitive to the DPS effects~\cite{Davies:2014rdz}. Specifically, when the DPS radiations contaminate the lepton's neighborhood, it leads to increased lepton-isolation values or failures in event topology selection, thereby reducing the lepton efficiency~\cite{Rehermann:2010vq,CMS:2015xaf,CMS:2018rym}. Consequently, standard analysis cuts preferentially reject events with high DPS activity, artificially suppressing the effective DPS fraction in the $Z$-boson calibration sample. In the $Z\to\ell\ell$ events, the selection criteria require two leptons, one more than in the $W\to\ell\nu$ events, and consequently veto more events with high DPS activity. The resulting parameterization of the $u_\parallel$ response and resolution, derived from this selection-biased dataset, therefore deviates from the true detector response in high-DPS environments. When the $Z$-derived $R$ is applied to $W$-boson analyses, the mismatch between the derived response function and the actual DPS conditions in $W$-boson events manifests as a non-closure in recoil-sensitive observables~\cite{Bizon:2019zgf}. Generally speaking, the $u_{\parallel}$ component is subject to various impacts from lepton reconstruction efficiencies, momentum scale calibrations, and selection biases~\cite{Davies:2014rdz}. Consequently, the DPS signal is heavily filtered by the sequential analysis selections, leaving only residual, highly cut-dependent indirect effects. Compared to $u_\parallel$, $u_\perp$ is much more sensitive to the DPS effects~\cite{Herget:2018uam}. Oriented perpendicular to the lepton direction, the $u_{\perp}$ component is largely insensitive to lepton momentum scale calibrations and isolation cuts. The isotropic semi-hard activity generated by DPS contributes directly to $u_{\perp}$, yielding a response that is direct and readily observable. However, $u_\perp$ is not included in the CDF calibration~\cite{CDF:2022hxs}, and the DPS effects on $u_\perp$ could be inadvertently neglected, thereby obscuring the DPS contributions in the $Z$-boson production and rendering the $Z$-derived $R$ insensitive to the DPS contributions. When applied to tune MC simulations, the $Z$-derived recoil response, which excludes the $u_{\perp}$-constraint, can deviate from the true detector response in high-DPS environments, preventing its direct transfer to $W$-boson analyses.

Noticeably, if additional QCD radiation merely rescales the hadronic recoil magnitude $u_{\mathrm{T}}=|\vec{u}_{\mathrm{T}}|$ without altering its underlying kinematic structure, the transfer of the $Z$-derived calibration $R$ to $W$-boson analyses remains acceptable, at least partially. However, in the DPS events, an extra independent parton scattering occurs in the same hadron-hadron collision. The hadronic activity from this second scattering is not necessarily correlated with the primary $W$-boson recoil. Thus, it randomly contributes to the hadronic recoil $\vec{u}_{\mathrm{T}}$ and introduces structural changes that go beyond simple rescaling: it alters the recoil composition, modifies the jet topology, and disrupts the correlation between visible hadronic activity and missing transverse momentum~\cite{CMS:2024lrd}. Additionally, these effects could induce extra sources of missing transverse momentum, for example through DPS-induced heavy-flavor semi-leptonic decays that contribute additional neutrinos uncorrelated with the primary $W$ decay. These $W$-specific effects break the process-independence of the detector response within matched kinematic bins, meaning the $Z$ recoil calibration cannot be directly transferred without additional modeling~\cite{Herget:2018uam}. Therefore, when the hadronic recoil $\vec{u}_{\mathrm{T}}$ is calibrated solely using the lepton-derived $\vec{p}_{\mathrm{T}}^{~Z}$, and then directly applied to $W$-boson events by replacing $\vec{p}_{\mathrm{T}}^{~Z}$ with $\vec{p}_{\mathrm{T}}^{~W}$, the DPS-induced activity may be incorrectly absorbed into the recoil model. Since the reconstructed missing transverse momentum in $W$-boson events depends directly on the calibrated recoil, i.e., $\vec{p}_{\mathrm{T}}^{~\mbox{\scriptsize miss}}=-(\vec{u}_{\mathrm{T}}+\vec{p}_{\mathrm{T}}^{~\ell\ell})$, such a misattribution will bias the transverse-mass distribution and hence the extracted $W$-boson mass. Therefore, the DPS effects present in $W$-boson events cannot be fully accounted for by the $Z$-derived recoil calibration, necessitating dedicated modeling, validation, or the inclusion of an additional systematic uncertainty.

On the other hand, the calibration of hadronic recoil using $Z\to\ell\ell$ events intrinsically depicts the detector response to \textit{visible} hadronic activity, rather than directly measuring the invisible transverse momentum $\vec{p}_{\mathrm{T}}^{~\mbox{\scriptsize miss}}$ itself. In the $Z$-boson decays, the high-precision measurements of both charged leptons allows the true boson transverse momentum to be reconstructed with negligible uncertainty, enabling the use of the transverse momentum balance condition $\vec{p}_{\mathrm{T}}^{~Z}+\vec{u}_{\mathrm{T}}\approx0$ to derive accurate response and resolution functions for the hadronic recoil $\vec{u}_{\mathrm{T}}$. In contrast, for the $W\to\ell\nu$ events, the missing transverse momentum $\vec{p}_{\mathrm{T}}^{~\mbox{\scriptsize miss}}=-(\vec{u}_{\mathrm{T}}+\vec{p}_{\mathrm{T}}^{~\ell})\approx\vec{p}_{\mathrm{T}}^{~\nu}$ is not directly measured but inferred from the visible system. Consequently, the reconstruction fidelity of $\vec{p}_{\mathrm{T}}^{~\mbox{\scriptsize miss}}$ and the derived transverse mass $m_{\mathrm{T}}$ remains critically dependent on how faithfully the calibrated recoil model $R$ tracks the true momentum balance in the presence of the undetected neutrino~\cite{Herget:2018uam}. Due to the undetectable neutrino, the $W$-boson $\vec{p}_{\mathrm{T}}^{~W}$ cannot be precisely reconstructed from data, even when $\vec{u}_{\mathrm{T}}$ is directly measured. This prevents a direct event-by-event correlation among $\vec{p}_{\mathrm{T}}^{~W}$, the charged lepton momentum $\vec{p}_{\mathrm{T}}^{~\ell}$, and $\vec{u}_{\mathrm{T}}$. Subsequently, it inherently relies on a chain of modeling assumptions~\cite{foppiani_2017_50tth-1x736}, including the generator-level description of DPS and UE, the selection of associated MC tunes, the parameterization of the recoil response, and so on. This is inevitably sensitive to DPS-like effects. Consequently, the calibration transfer must rely on the generator-level $\vec{p}_{\mathrm{T}}^{~W}$ as an input, introducing a significant discrepancy in the truth-to-detector mapping. Both the fidelity of the $W$-event detector simulation and the robustness of the calibration procedure contribute to this discrepancy, thereby introducing DPS-dependent systematic uncertainties into the extracted $M_W$ value. To ensure the reliability of the $\vec{p}_{\mathrm{T}}^{~\mbox{\scriptsize miss}}$ measurement, it is advisable to incorporate ``directly measured hadronic information''~\cite{ATLAS:2018txj}; see section~\ref{sec:4} for more details.

\begin{figure}
\centerline{
\includegraphics[width=\columnwidth]{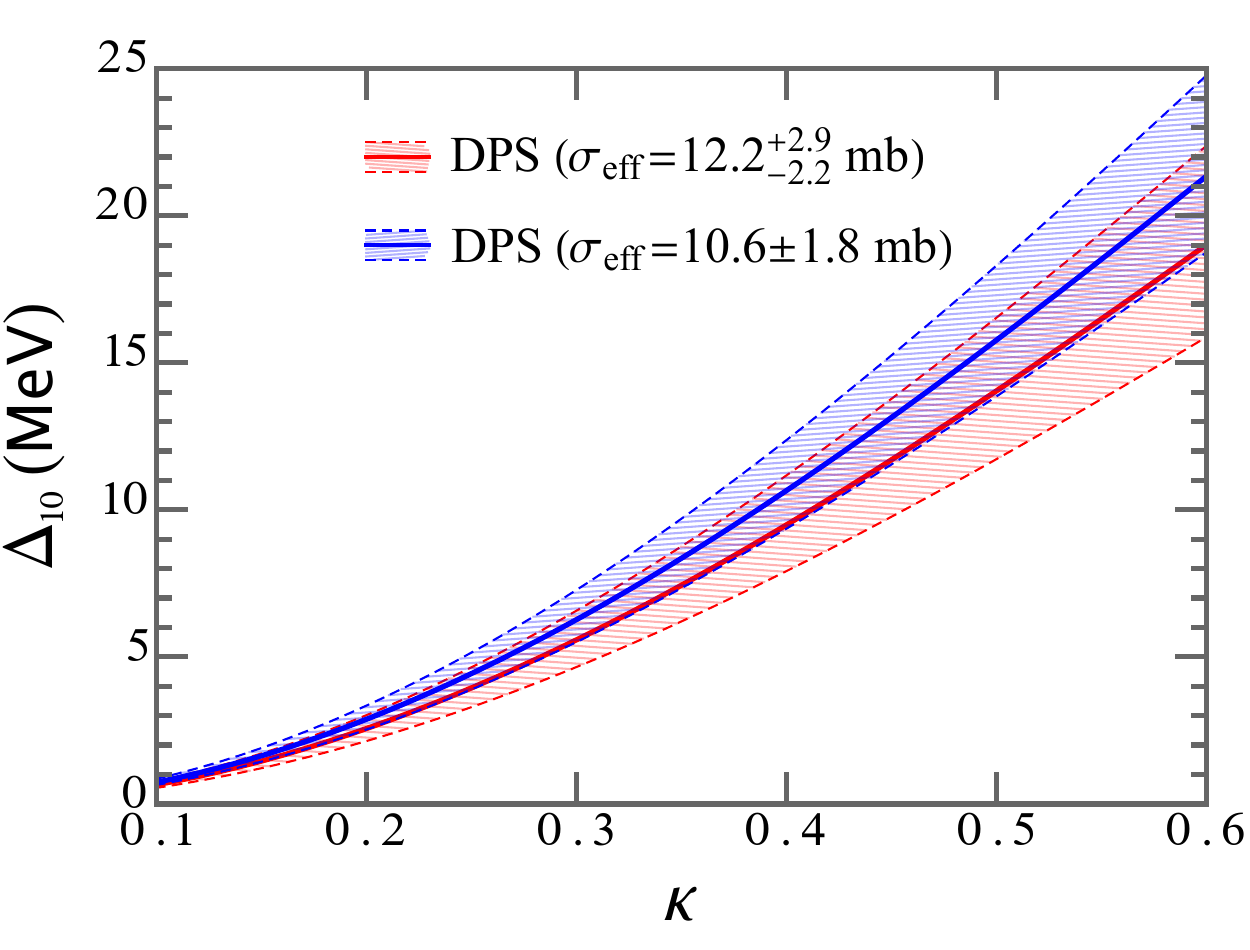}
}
\caption{The shift parameter $\Delta_{10}$, defined in equation~\eqref{MWfit0}, as a function of $\kappa$, extracted from the fits to the $W$-boson events, with all settings and conventions as in figure~\ref{fig:2}. 
}
\label{fig:4}
\end{figure}

For completeness, we assess the residual DPS effects on the CDF $M_W$ measurement under some theoretical assumptions. 
As noted above, the DPS contributions could be partially lost in the $Z$-based CDF calibration;
for instance, the $Z$-derived $R$ is intrinsically a $u_\parallel$ response, which is relatively insensitive to the DPS activity, 
and the $u_\perp$ data is not included in the calibration, thereby obscuring the DPS contributions in the $Z$-boson production. 
If the DPS effects are already partially accounted for in the hadronic recoil calibration, we could assume that a fraction $\kappa$ of such DPS contributions in the spectator missing transverse momentum is effectively identified and incorporated into the $u_{\parallel}$-based calibration. 
Generally, according to equation~\eqref{eq:p_DPS^miss}, the missing transverse momentum can be approximately expressed as
\begin{align}
\label{eq:pTmissDPS1}
p_{\mathrm{T},\mbox{\scriptsize DPS}}^{\mbox{\scriptsize miss}}\approx p_{\mathrm{T},\mbox{\scriptsize SPS}}^{\mbox{\scriptsize miss}}+p_{\mathrm{T},jj}^{\mbox{\scriptsize miss}}\cos\alpha+\frac{\left(p_{\mathrm{T},jj}^{\mbox{\scriptsize miss}}\right)^2}{2p_{\mathrm{T},\mbox{\scriptsize SPS}}^{\mbox{\scriptsize miss}}}\left(1-\cos^2\alpha\right)\,. 
\end{align}
where the second term is $\sim4$~GeV, whereas the third term is $\sim0.1$~GeV. 
The structure of the second term actually suggests an approach to quantify the calibration-induced DPS effects. 
Assuming that a fraction $\kappa$ of the DPS contributions in $W$ boson production can be accounted for by the $Z$-derived calibration $R$, we have
\begin{align}
p_{\mathrm{T},\mbox{\scriptsize DPS}}^{\mbox{\scriptsize miss}}\approx p_{\mathrm{T},\mbox{\scriptsize SPS}}^{\mbox{\scriptsize miss}}+\kappa\times p_{\mathrm{T},jj}^{\mbox{\scriptsize miss}}\cos\alpha\,, 
\end{align}
where the third term in equation~\eqref{eq:pTmissDPS1} is neglected due to its subleading kinematic contribution. 
Since $u_\parallel$ is typically less sensitive to the DPS effects than $u_\perp$, the former can therefore be expected to carry a smaller fraction of DPS contributions than the latter, thereby contributing no more than half of the total, particularly when transferring the $Z$-derived $R$ to the $W$-boson events. Given that the CDF calibration relies solely on the $u_{\parallel}$ component, $\kappa$ is expected to be $\lesssim0.5$. 
Then, the $p_{\mathrm{T},\mbox{\scriptsize DPS}}^{\mbox{\scriptsize miss}}$ distribution becomes
\begin{align}
\frac{\mathrm{d}\sigma^{\mbox{\scriptsize{DPS}}}}{\sigma^{\mbox{\scriptsize{DPS}}}\mathrm{d} p_{\mathrm{T},\mbox{\scriptsize DPS}}^{\mbox{\scriptsize miss}}}=\int_0^\pi\frac{\mathrm{d}\alpha}{\pi}\int B(p_{\mathrm{T},\mbox{\scriptsize DPS}}^{\mbox{\scriptsize miss}}-\kappa\times p_{\mathrm{T},jj}^{\mbox{\scriptsize miss}}\cos\alpha)D(p_{\mathrm{T},jj}^{\mbox{\scriptsize miss}})\mathrm{d} p_{\mathrm{T},jj}^{\mbox{\scriptsize miss}}\,. 
\end{align}
In addition, we assume that DPS does not significantly alter the direction of the missing transverse momentum, as its contribution is subleading. 
Accordingly, by the definition of $m_{\mathrm{T}}\equiv\sqrt{2\left(p_{\mathrm{T}}^\ell p_{\mathrm{T}}^{\mbox{\scriptsize miss}}-\vec{p}_{\mathrm{T}}^{~\ell} \cdot \vec{p}_{\mathrm{T}}^{~\mbox{\scriptsize miss}}\right)}$, the corresponding transverse mass is modified to
\begin{align}
m_{\mathrm{T},\mbox{\scriptsize DPS}}=m_{\mathrm{T},\mbox{\scriptsize SPS}}\sqrt{\frac{p_{\mathrm{T},\mbox{\scriptsize DPS}}^{\mbox{\scriptsize miss}}}{p_{\mathrm{T},\mbox{\scriptsize SPS}}^{\mbox{\scriptsize miss}}}}\approx m_{\mathrm{T},\mbox{\scriptsize SPS}}\sqrt{1+\kappa\times \frac{p_{\mathrm{T},jj}^{\mbox{\scriptsize miss}}\cos\alpha}{p_{\mathrm{T},\mbox{\scriptsize SPS}}^{\mbox{\scriptsize miss}}}}\,, 
\end{align}
yielding the normalized differential distribution
\begin{align}
\frac{\mathrm{d}\sigma^{\mbox{\scriptsize{DPS}}}}{\sigma^{\mbox{\scriptsize{DPS}}}\mathrm{d} m_{\mathrm{T},\mbox{\scriptsize DPS}}}=&\int B(p_{\mathrm{T},\mbox{\scriptsize SPS}}^{\nu})\mathrm{d} p_{\mathrm{T},\mbox{\scriptsize SPS}}^{\nu} D(p_{\mathrm{T},jj}^{\nu})\mathrm{d} p_{\mathrm{T},jj}^{\nu}\nonumber\\
&\!\!\!\!\!\!\!\!\!\!\!\!\!\!\times\int_0^\pi\frac{\mathrm{d}\beta}{\pi}C\left(\frac{m_{\mathrm{T},\mbox{\scriptsize DPS}}}{\sqrt{1+\kappa\times \frac{p_{\mathrm{T},jj}^{\mbox{\scriptsize miss}}\cos\alpha}{p_{\mathrm{T},\mbox{\scriptsize SPS}}^{\mbox{\scriptsize miss}}}}}\right)\Bigg{/}\sqrt{1+\kappa\times \frac{p_{\mathrm{T},jj}^{\mbox{\scriptsize miss}}\cos\alpha}{p_{\mathrm{T},\mbox{\scriptsize SPS}}^{\mbox{\scriptsize miss}}}}\,. 
\end{align}
As shown, $\kappa$ parameterizes the performance of the hadronic recoil calibration in accounting for the DPS effects on the $p_{\mathrm{T}}^{\mbox{\scriptsize miss}}$ and $m_{\mathrm{T}}$ distributions. 
In principle, it should include all DPS contributions, such as those from the $u_\parallel$ and $u_\perp$ components. In practice, however, the current CDF calibration mainly captures the $u_\parallel$ component. 
Following the procedure outlined in the main text, we also perform simulations and fits for several benchmark $\kappa$ values. In particular, $\Delta_{10}$ can be treated as a function of $\kappa$, where $\Delta_{10}$ is defined as in equation~\eqref{MWfit0}. Then, we find the calibration-derived DPS effects on $M_W$ to be $\Delta_{10}=14^{+2}_{-2}\times\left(\kappa/0.5\right)^{1.77}~\mbox{MeV}$ and $\Delta_{10}=16^{+2}_{-2}\times\left(\kappa/0.5\right)^{1.77}~\mbox{MeV}$ for $\sigma_{\mbox{\footnotesize eff}}=12.2^{+2.9}_{-2.2}~\mbox{mb}$~\cite{CMS:2022pio} and $\sigma_{\mbox{\footnotesize eff}}=10.6\pm1.8~\mbox{mb}$~\cite{ATLAS:2025bcb}, respectively. More details can be found in figure~\ref{fig:4}. Note that in equation~\eqref{MWfit0}, $\Delta_{10}\sim60$~MeV. 
Thus, in the $M_{W}$ measurement, the DPS effects captured by the $u_{\parallel}$-based CDF calibration amount to only $\sim25\%$ of the total DPS contribution for $\kappa\sim0.5$. Now it is safe to conclude that if $\kappa$ is less than 0.5, at least $\sim75\%$ of the total DPS effects on the $W$-boson mass remain unmitigated by the $Z$-based CDF calibration. 

Nevertheless, it is still unclear to what extent DPS contributions in the $W$ boson production can be accounted for by the current CDF calibration, meaning $\kappa$ is poorly constrained. Additionally, the approach presented here relies heavily on our theoretical assumptions. 
This will serve as a source of uncertainty in our estimate. Notably, a full assessment of these assumptions and their associated limitations requires a dedicated CDF experimental analysis, specifically incorporating DPS MC modeling alongside N$^3$LL+NNLO QCD calculations. Furthermore, another key challenge lies in whether the derived quantities, such as $\vec{p}_{\mathrm{T}}^{~\mbox{\scriptsize miss}}$ and $m_{\mathrm{T}}$, can be reliably obtained and corrected when transferring a calibration derived from one process to another. Also there are other challenges discussed above that prevent the CDF calibration from fully capturing DPS effects, consequently introducing a bias in the $M_W$ measurement.

As mentioned above, in the current CDF analysis, the recoil response function $R$ is calibrated against $\vec{p}_{\mathrm{T}}^{~Z}$~\cite{CDF:2022hxs}, thereby misidentifying the extra energy from DPS radiation as the $W$ recoil $\vec{u}_{\mathrm{T}}^{~W}$. Consequently, the CDF calibration cannot fully account for the DPS contributions, which inevitably leads to a bias in the measured $W$-boson mass. In future CDF analyses, we should incorporate ``directly measured hadronic information''; see section~\ref{sec:4} for more details. This implies that the calibration should not rely solely on the theoretical balance value derived from leptons, but must also account for the distribution, scale, resolution, and residuals of the measured hadronic recoil $\vec{u}$ itself. After including ``directly measured hadronic information'', the $W$-like approach will enable more realistic modeling of effects such as DPS and additional radiation~\cite{ATLAS:2018txj,Beguin:2019bex}.

\end{document}